\documentclass[epj]{svjour}
\usepackage{times}
\usepackage{graphics}
\usepackage{amsmath}
\usepackage{amsfonts}
\usepackage{amssymb}
\usepackage{dsfont}
\usepackage{graphicx}
\usepackage{slashed}
\usepackage{yfonts}
\usepackage{nicefrac}
\usepackage{multirow}
\usepackage{widetext}
\usepackage[usenames]{color}
\usepackage[colorlinks=true,linkcolor=blue,citecolor=blue,urlcolor=blue]{
hyperref}
\usepackage{cite}
\usepackage{dsfont}

\usepackage{amsmath}
\usepackage{amssymb}
\usepackage{graphicx}
\usepackage{dsfont}
\usepackage{dcolumn}
\usepackage{units}
\usepackage{wasysym}
\usepackage{multirow}
\usepackage{slashed}
\usepackage[usenames]{color}
\usepackage{slashed}
\usepackage{booktabs}
\usepackage{wasysym}

\newcommand{\Slash}[1]{\ooalign{\hfil/\hfil\crcr$#1$}}
\newcommand{\Eq}[1]{(\ref{#1})}

\definecolor{violet}{RGB}{111,0,255}
\definecolor{lred}{rgb}{1,0.90,0.7}
\definecolor{dgreen}{rgb}{0.0,0.50,0.0}
\definecolor{dblue}{rgb}{0,0,0.5}
\begin{document}
\title{Electromagnetic transition form factors of baryons in the 
space-like momentum region}

\author{H\`elios Sanchis-Alepuz\inst{1}, Reinhard Alkofer\inst{1} and Christian S. Fischer\inst{2}    
}                     
\institute{Institut f\"ur Physik, Karl-Franzens--Universit\"at 
Graz, NAWI Graz, 8010 Graz, Austria\\ \and
Institut f\"ur Theoretische Physik, Justus-Liebig--Universit\"at 
Giessen, 35392 Giessen, Germany }
\date{Received: date / Revised version: date}
\abstract{We present results from a calculation of the electromagnetic transition form factors between ground-state octet 
and decuplet baryons as well as the octet-only $\Sigma^0$ to $\Lambda$ transition. We work in the combined framework of 
Dyson-Schwinger equations and covariant Bethe-Salpeter equations with all elements, the baryon three body wave function,
the quark propagators and the dressed quark-photon vertex determined from a well-established, momentum dependent approximation
for the quark-gluon interaction. We discuss in particular the similarities among the different transitions as well as the 
differences induced by $SU(3)$-isospin symmetry breaking. We furthermore provide estimates for the slopes of the electric and
magnetic $\Sigma^0$ to $\Lambda$ transitions at the zero photon momentum point. 
 % end of PACS codes
} %end of abstract
\titlerunning{Electromagnetic transition form factors of baryons}
\maketitle

\section{Introduction}
\label{sec:introduction}

The spectrum and the structure of hadrons are an important experimental handle to understand the formation of observable matter 
from quarks and gluons, the elementary degrees of freedom of quantum chromodynamics(QCD). The internal structure of hadrons is 
particularly sensitive to the details of the binding interaction. Form factors provide information on this internal structure. 
Electromagnetic form factors, in particular, encode how electric charge and multipole moments as well as magnetic multipole moments
are distributed inside the hadron. 
Moreover, form factors in the spacelike photon-momentum region govern the electromagnetic interaction of hadrons with other 
charged particles via the exchange of virtual photons.

Currently, in the baryon sector most of the experimental knowledge on form factors in the spacelike region comes from experiments 
on meson photo- and electro-production off nucleons. In recent years, experimental data gathered chiefly from experiments at 
MIT-Bates, Jefferson Lab and MAMI (see, e.g. \cite{Bernauer:2013tpr,Punjabi:2015bba} for reviews of experimental data) have 
provided a rather precise picture of the spacelike neutron and proton elastic form factors, some of its features still challenging 
theoretical explanations. 

Transition form factors of baryons provide additional information. In the case of nucleons, the electromagnetic nucleon to Delta 
transition allows to study, for example, the possible deformation of charge and magnetisation distributions, otherwise inaccessible 
for spin-$\nicefrac{1}{2}$ baryons. Transition form factors are also a necessary input for any calculation of the production processes 
studied in scattering experiments. In this respect the main electromagnetic transition of interest is between the nucleon and any 
of its non-strange excitations (chiefly the $\Delta(1232)$) in the s-channel. For this transition, there are abundant experimental 
data as well (see, e.g. \cite{Aznauryan:2012ba}; electromagnetic transitions between hyperons play also a relevant role in the u-channel).

The study of hyperon form factors, in addition, allows to understand the role of $SU(3)$ flavour-symmetry breaking in the properties 
of hadrons. Unfortunately, experimental knowledge of the structure of hyperons is limited to some values for static properties \cite{Olive:2016xmw},
to the measurement of elastic form factors at a small number of large time-like photon momenta \cite{Dobbs:2014ifa} and the branching 
ratios of some decays \cite{Keller:2011nt,Keller:2011aw}. The study of the hyperon structure for spacelike photon momenta is also one 
of the physics goals of the CLAS $12$-GeV upgrade \cite{Dudek:2012vr}. At small timelike momenta, the future PANDA experiment at the 
Facility for Antiproton and ion Research (FAIR) offers the interesting possibility to explore the $\Sigma^0$-$\Lambda$ transition form 
factor by the Dalitz decay $\Sigma^0 \rightarrow \Lambda e^+ e^-$. An important task for theory is then to provide for a precise calculation
of both, the hadronic contributions but also the electromagnetic corrections to this decay in order to be able to unambiguously extract 
the hadronic part at small time-like momenta up to energies of $\Sigma^0$-$\Lambda$, see \cite{Granados:2017cib} for a detailed discussion.  

This work continues the study of the internal properties of baryons in the combined Dyson-Schwinger (DSE) and Bethe-Salpeter (BSE) 
framework using three-body equations \cite{Eichmann:2009qa,Eichmann:2011vu,Eichmann:2011pv,Sanchis-Alepuz:2013iia,Sanchis-Alepuz:2015fcg,Eichmann:2016hgl}. In principle,
this approach provides a direct connection between QCD as a quantum field theory and hadronic observables. The dynamics of quarks and 
gluons as constituents of bound states, as well as the interaction vertices among them are determined by the infinite set of coupled DSEs; 
these elements then define the relevant BSEs, with solutions for amplitudes encoding all the properties of hadrons as bound states of 
quarks and gluons. In practice, for most applications, both the DSEs and the BSEs are truncated in a consistent manner, such that all 
the relevant symmetries (in particular chiral symmetry) are preserved; see, e.g. \cite{Eichmann:2016yit} for a review of the formalism
and a detailed discussion of truncation schemes. 

This paper is organised as follows. In section~\ref{sec:formalism} we give a brief overview of the DSE/BSE formalism in order to keep 
this work reasonably self-contained. In section \ref{sec:results} we discuss the results of the electromagnetic transitions from the 
baryon decuplet to the baryon octet, stressing the similarities and the differences with the well-studied nucleon to Delta transition. 
Additionally we have calculated the form factors for the octet-only $\Sigma^0$-$\Lambda$ transition. We summarise and conclude in 
section \ref{sec:summary}. The appendices contain some technical details of the calculation.
 
%%%%%%%%%%%%%%%%%%%%%%%%%%%%%%%%%%%%%%%%%%%%%%%%%%%%%%%%%%%%%%%%%%%%%
%                        S E C T I O N                              %
%%%%%%%%%%%%%%%%%%%%%%%%%%%%%%%%%%%%%%%%%%%%%%%%%%%%%%%%%%%%%%%%%%%%%

\section{Short summary of the formalism}\label{sec:formalism} 

\subsection{Bound state equations }

\begin{figure*}[hbtp]
 \begin{center}
  \includegraphics[width=0.8\textwidth,clip]{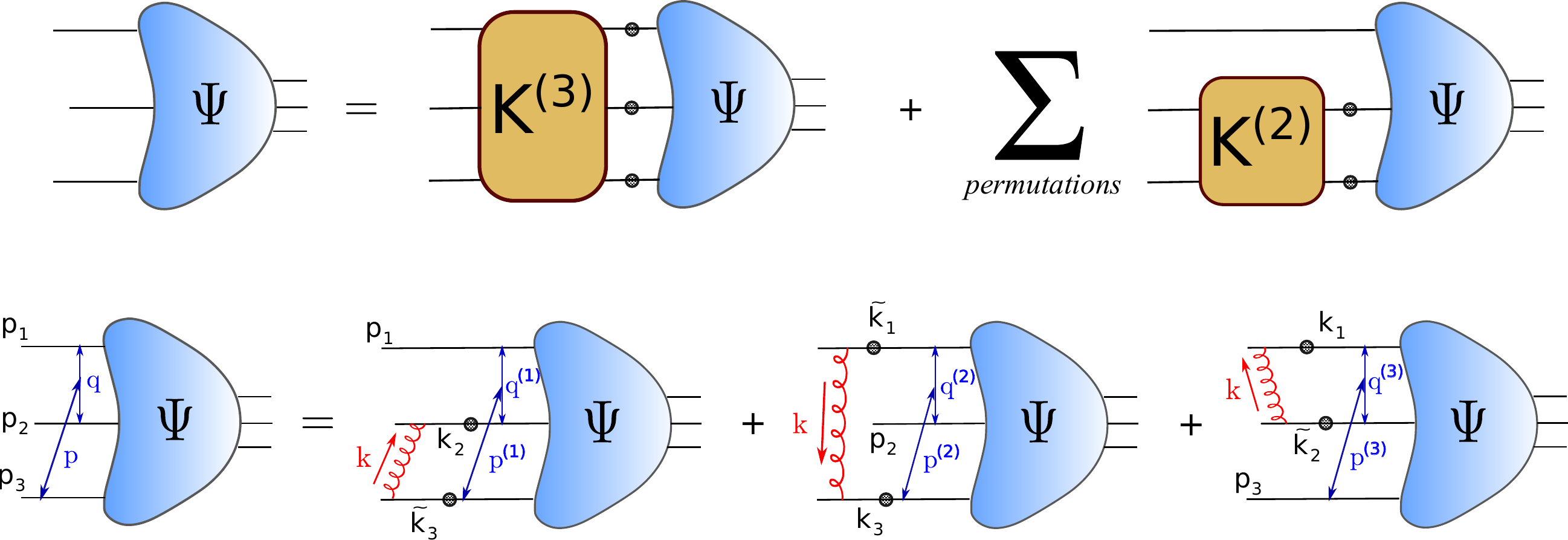}
 \end{center}
 \caption{Upper figure: Diagrammatic representation of the three-body Bethe-Salpeter 
equation with amplitude $\Psi$. Full quark propagators are denoted by straight lines with black dots.
The interaction between the quarks are encoded in the three-body and two-body
kernels $K^{(3)}$ and $K^{(2)}$.
Lower figure: Faddeev equation in the rainbow-ladder 
truncation; gluon lines are dressed with the effective coupling, Eq.~(\ref{eq:MTmodel}). 
See Appendix~\ref{sec:kinematics} for the definitions of momenta.}\label{fig:faddeev_eq}
\end{figure*}

The combined DSE/BSE framework for baryons has been described in full detail in several references (see, e.g. \cite{Eichmann:2016yit,Sanchis-Alepuz:2015fcg,Eichmann:2011vu}). In order to keep this article reasonably self-contained, 
we recapitulate here only its key elements.

In the DSE/BSE formalism, baryons are described as bound states of three quarks, and their properties are obtained as solutions of a three-body Bethe-Salpeter equation. The Bethe-Salpeter amplitude is a tensor product of colour, flavour and spin-momentum parts
\begin{equation}\label{eq:BSE_amplitude}
 \Gamma_{ABCD}(p,q,P)=\left(\sum_\rho 
\Psi^\rho_{\alpha\beta\gamma\mathcal{I}}(p,q,P) \otimes 
F^\rho_{abcd}\right)\otimes \frac{\epsilon_{rst}}{\sqrt{6}}~.
\end{equation}
where we use $\{ABC\}$ as collective spin, flavour and colour indices for the valence quarks and $D$ 
for the resulting baryon. This amplitude is the solution of the 
three-body Bethe-Salpeter equation Fig.~\ref{fig:faddeev_eq}.
The amplitude depends on the three quark momenta $p_{1,2,3}$, which can be 
expressed in terms of two relative momenta $p$ and $q$ and the total momentum 
$P$ (see Eq.~\Eq{eq:defpq} in Appendix~\ref{sec:kinematics}).
The colour term $\epsilon_{rst}/\sqrt{6}$ fixes the baryon to be a colour singlet 
and the flavour terms $F^\rho_{abcd}$ are the quark-model $SU(3)$ irreducible representations.
The spin-momentum part of the Bethe-Salpeter amplitude, 
$\Psi^\rho_{\alpha\beta\gamma\mathcal{I}}(p,q,P)$, is a tensor with three Dirac 
indices $\alpha,\beta,\gamma$ associated to the valence quarks and a generic 
index $\mathcal{I}$ which depends on the spin of the resulting bound 
state. One then introduces a covariant basis $\{\tau(p,q;P)\}$ and expands the spin-momentum term
\begin{equation}\label{eq:basis_expansion}
\Psi^\rho_{\alpha\beta\gamma\mathcal{I}}(p,q,P)=\sum_i f^{\rho}_i(p^2,q^2,z_0,z_1,
z_2)\tau_{\alpha\beta\gamma\mathcal{I}}^i(p,q;P)~,
\end{equation}
where the scalar coefficients $\{f\}$ depend on Lorentz scalars $p^2$, $q^2$, 
$z_0=\widehat{p_T}\cdot\widehat{q_T}$, $z_1=\widehat{p}\cdot\widehat{P}$ and 
$z_2=\widehat{q}\cdot\widehat{P}$ only. The subscript $T$ denotes transverse 
projection with respect to the total momentum and vectors with hat are 
normalised. 

The resulting Bethe-Salpeter equation (with the three-body irreducible term neglected, see below) in its final form reads
\begin{flalign}\label{eq:Faddeev_coeff}
 f^{\rho}_i(p^2,q^2,z_0,z_1,z_2)=& \nonumber\\        
C\mathcal{F}_1^{\rho\rho';\lambda}~H_1^{ij}&~g^{\rho'',\lambda}_{j}(p'^2,q'^2,
z'_0,z'_1,z'_2)+ \nonumber\\
C\mathcal{F}_2^{\rho\rho';\lambda}~H_2^{ij}&~g^{\rho'',\lambda}_{j}(p''^2,q''^2,
z''_0,z''_1,z''_2)+ \nonumber\\
C\mathcal{F}_3^{\rho\rho';\lambda}~&~g^{\rho',\lambda}_{i}(p^2,q^2,z_0,z_1,z_2)~,
\end{flalign}
with
%\begin{widetext}
\begin{flalign}\label{eq:Faddeev_coeff3}
g^{\rho,\lambda}_i(p^2,q^2,z_0,z_1,z_2)= &\nonumber\\
\int_k~\Bigl[~\bar{\tau}^
{i}_{\beta\alpha\mathcal{I}\gamma}(p,q,P)& K_{\alpha\alpha',\beta\beta'}(p,q,
k) \delta_{\gamma\gamma''} \times  \nonumber\\ 
S^{(\lambda_1)}_{\alpha'\alpha''}(k_1) S^{(\lambda_2)
}_{\beta'\beta''}&(k_2)~\tau^{j}_{\alpha''\beta''\gamma''\mathcal{I}}(p_{(3)},q_{
(3)},P)~\Bigr]\times \nonumber\\ 
&f^{\rho}_j(p_{(3)}^2,q_{(3)}^2,z^{(3)}_0,z^{(3)}_1,z^{(3)}_2)~.
\end{flalign}
%\end{widetext}
Here the factor $C=-\nicefrac{2}{3}$ stems from from colour traces. The flavour matrices $\mathcal{F}$, 
the rotation matrices $H_{1,2}$ and the different momenta are defined in Appendix \ref{sec:kinematics}.
The dressed quark propagators $S$ and the Bethe-Salpeter kernels $K$ are discussed below.

Covariant bases for the Dirac structure of the Bethe-Salpeter amplitudes (\ref{eq:basis_expansion}) 
can be obtained using symmetry requirements only; 
for positive-parity spin-$\nicefrac{1}{2}$ baryons it contains 64 elements 
\cite{Eichmann:2009en,Eichmann:2009zx} whereas for spin-$\nicefrac{3}{2}$ 
baryons it contains 128 elements \cite{SanchisAlepuz:2011jn}. In this way, 
one only needs to solve for the scalar functions $f$.
The index $\lambda$ runs over all elements in a given flavour state and the quark at the position 
$\ell$ in each term of the flavour wave 
function is denoted by the superindex $\lambda_\ell$ to keep track of the different flavours. The internal relative 
momenta $p_{(3)}$, $q_{(3)}$ (and analogously for $z^{(3)}_0$, $z^{(3)}_1$ and 
$z^{(3)}_2$)  are defined in Appendix \ref{sec:kinematics}. The conjugate 
of the covariant basis $\bar{\tau}$ has been defined in 
\cite{Eichmann:2009en,SanchisAlepuz:2011jn} and it is assumed that the basis 
$\{\tau\}$ is orthonormal.

The interaction kernel $K$ in the most general three-body Bethe-Salpeter equation (see Fig.~\ref{fig:faddeev_eq} consists of an irreducible three-body part and (permutations of)
two-body interactions
\begin{flalign}
K = K^{(3)} + \sum_a S^{-1}_a K^{(2)}_a
\end{flalign}
In the Faddeev approximation the three-body part $K^{(3)}$ is neglected, and we refer to the simplified BSE as the 
Faddeev equation (FE). In this work we truncate the two-body kernel $K^{(2)}$ to a ladder kernel, 
which consists of a single dressed gluon-exchange with vector coupling to the quark legs:
\begin{flalign}\label{eq:RLkernel}
K^{(2)}_a = Z_2^2 \frac{4\pi\alpha_{eff}(k^2)}{k^2}T^{\mu\nu}(k)\gamma_\mu \gamma_\nu
\end{flalign}
with $T^{\mu\nu}(k)=\delta^{\mu\nu}-\hat{k}^\mu \hat{k}^\nu$ the transverse projector.
This interaction is then specified by an effective coupling \cite{Maris:1997tm,Maris:1999nt} which has been employed 
frequently in hadron studies within the rainbow-ladder BSE/DSE framework. 
This model performs very well for phenomenological calculations of ground- and excited-state meson and baryon properties in selected channels
including the octet and decuplet baryons studied in this work \cite{Eichmann:2016yit,Eichmann:2016hgl}. It is defined as
\begin{flalign}\label{eq:MTmodel}
\alpha_{\textrm{eff}}(k^2) {}=&
 \pi\eta^7\left(\frac{k^2}{\Lambda^2}\right)^2
e^{-\eta^2\frac{k^2}{\Lambda^2}}\nonumber\\ &+{}\frac{2\pi\gamma_m
\big(1-e^{-k^2/\Lambda_{t}^2}\big)}{\textnormal{ln}[e^2-1+(1+k^2/\Lambda_{QCD}
^2)^2]}\,. 
\end{flalign}
with $k$ the momentum of the exchanged gluon. This interaction reproduces the one-loop QCD behaviour of the quark propagator at 
large momenta and the Gaussian distribution of interaction strength in the intermediate 
momentum region provides enough strength for dynamical chiral symmetry breaking to take place. 
The scale $\Lambda_t=1$~GeV is introduced for technical reasons and has no impact on
the results. For the anomalous dimension we use $\gamma_m=12/(11N_C-2N_f)=12/25$,
corresponding to $N_f=4$ flavours and $N_c=3$ colours. The scale in the ultraviolet 
part of the coupling is set to $\Lambda_{QCD}=0.234$ GeV.
In the infrared momentum region, the interaction strength is characterised by a 
scale $\Lambda$ and a dimensionless parameter $\eta$ that controls the width of 
the interaction. 
The scale $\Lambda=0.72$~GeV is adjusted to reproduce the 
experimental pion decay constant from the truncated pion BSE.
This as well as many other pseudo-scalar ground-state observables, such as the masses of 
ground-state mesons and baryons, turn out to 
be almost insensitive to the value of $\eta$ in the range of values of $\eta$ 
between $1.6$ and $2.0$ see, e.g. 
\cite{Krassnigg:2009zh,Nicmorus:2010mc,Eichmann:2011vu}). 
The $u/d$ and $s$ current-quark masses are 
fixed to reproduce the physical pion and kaon masses, respectively. 
The corresponding values are $m_{u/d}(\mu^2)=3.7$~MeV and 
$m_s(\mu^2)=85$~MeV. The renormalisation scale is chosen to be $\mu^2= (19 
\,\mbox{GeV})^2$.

The remaining necessary element in Eq.\eqref{eq:Faddeev_coeff3} is the full quark propagator 
$S$ (omitting now Dirac indices)
for the quark fla\-vours of interest. These are obtained as solutions of 
the quark DSE 
\begin{equation}\label{eq:quarkDSE}
 S^{-1}(p)=S^{-1}_{0}(p)+Z_{1f}g^2 C_F \!\!\int_q \!\gamma^\mu
D_{\mu\nu}(p-q)\Gamma^\nu(p,q)S(q)\,\,,
\end{equation}
which also contains the full quark-gluon vertex $\Gamma^\nu$ and the full gluon 
propagator $D_{\mu\nu}$; 
$S_{0}$ is the (renormalised) bare propagator with inverse
\begin{equation}\label{eq:bare_prop}
 S_{0}^{-1}(p)=Z_2\left(i\Slash{p}+m\right)\,,
\end{equation}
where $m$ is the bare quark mass 
and $Z_{1f}$ and $Z_2$ are renormalisation constants and $g$ the renormalised strong 
coupling. This equation is truncated to its rainbow version, which amounts to the replacement
\begin{align}
Z_{1f} \frac{g^2}{4\pi} D_{\mu\nu}(k)\Gamma^\nu(p,q) = Z_2^2 T_{\mu\nu}(k) 
\frac{\alpha_{\mathrm{eff}}(k^2)}{k^2}\gamma^\nu\;,
\end{align}
with $T_{\mu\nu}(k)$ the transverse projector, {\it cf.}, eq.~(\ref{eq:RLkernel}). The renormalisation 
constants are chosen such that multiplicative renormalisability is preserved. 
In combination with a ladder-truncated meson BSE, this truncation also preserves chiral symmetry and its QCD 
breaking pattern ensuring the validity of the Gell-Mann-Oakes-Renner relation and the (pseudo-)Goldstone nature of the pion.

\subsection{Form factor calculation}\label{sec:gauging}

\begin{figure*}[hbtp]
 \begin{center}
  \includegraphics[width=0.8\textwidth,clip]{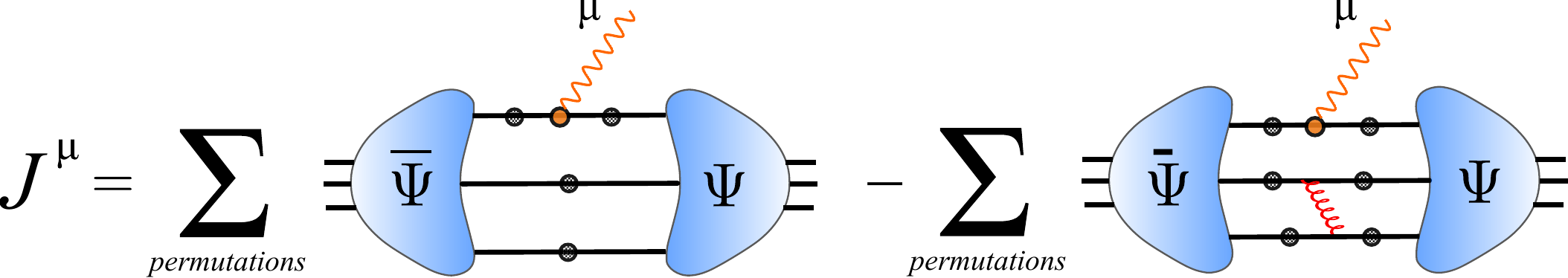}
 \end{center}
 \caption{Diagrams necessary to calculate the current $J^\mu$, describing the coupling of an external photon field to a baryon, 
 described by the Faddeev amplitude $\Psi$, in the RL truncation.}\label{fig:current_diagrams}
\end{figure*}

The procedure to couple an external field to a BSE equation is called \textit{gauging} of the equation and was introduced
in \cite{Haberzettl:1997jg,Kvinikhidze:1998xn,Kvinikhidze:1999xp,Oettel:1999gc,Oettel:2000jj}. 
The main features of this procedure is that it ensures gauge invariance in the coupling with an external electromagnetic field 
(hence charge conservation) and prevents the over-counting of diagrams.
The specific application of this formalism in the rain\-bow-ladder (RL) truncated
three-body BSE framework has been already described in \cite{Eichmann:2011vu}. 
We refrain from repeating the steps here and give only
the final expression, corresponding to the diagrams in Fig.~\ref{fig:current_diagrams}, with emphasis in their flavour dependence.

The form factors are extracted from the current $J^\mu$ describing the coupling of baryons to an external electromagnetic current which,in the rainbow-ladder truncation,
is
\begin{flalign}\label{eq:FFeqRL_simple}
J_{\mathcal{I}'\mathcal{I}}^\mu=\sum_{\rho\rho';\lambda}(&\mathcal{Q}^{\rho\rho'
}_1(\mathcal{F}_1^{\rho\rho'}J_3^{\rho'\rho'';\lambda}\mathcal{F}_1^{T,
\rho''\rho})_{ \mathcal{I}'\mathcal{I}}^\mu+\nonumber\\
&\mathcal{Q}^{\rho\rho'}_2(\mathcal{F}_2^{\rho\rho'}J_3^{\rho'\rho'';\lambda}
\mathcal{F}_2^{T,\rho''\rho})_{\mathcal{I}'\mathcal{I}}^\mu+\nonumber\\
&\mathcal{Q}^{\rho\rho'}_3(J_3^{\rho\rho';\lambda})_{\mathcal{I}'\mathcal{I}}
^\mu)~,
\end{flalign}
with
\begin{widetext}
\begin{flalign}\label{eq:FFeqRL}
(J_3^{\rho\rho';\lambda})_{\mathcal{I}'\mathcal{I}}^\mu&=\int_p\int_q\bar{\Psi}
^\rho_{\beta'\alpha'\mathcal{I}'\gamma'}(p^{\{3\}}_f,q^{\{3\}}
_f,P_f)\left[S^{(\lambda_1)}_{\alpha'\alpha}(p_1)S^{(\lambda_2)}_{\beta'\beta}
(p_2)\left(S^{(\lambda_3)}(p_3^f)\Gamma^\mu(p_3,Q)S^{(\lambda_3)}(p_3^i)\right)_{\gamma'\gamma}\right]
\times\nonumber\\
~~~~~~~~~~~~~~~~~~~~~~~~~~~~~~&\left(\Psi^{\rho'}_{\alpha\beta\gamma\mathcal{I}}(p^{\{3\}
}_i,q^{\{3\}}_i,P_i)-\Psi^{\{3\};\lambda}_{\alpha\beta\gamma\mathcal{I}}(p^{\{3 
\}}_i,q^{
\{3\}}_i,P_i)\right)~,
\end{flalign}
\end{widetext}
where we have defined $\Psi^{\{3\};\lambda}$ as the amplitude obtained from the result of the third term in the Faddeev equation \eqref{eq:Faddeev_coeff3}. The matrices $\mathcal{F}$ results from the flavour contractions for the different baryons, as given in Appendix~\ref{sec:flavour}.
The final and initial momenta of the interacting quark $\kappa$ are
%\begin{equation}
$
 p_\kappa^{\nicefrac{f}{i}}=p_\kappa\pm\frac{Q}{2}~,
$
%\end{equation}
with $Q=P_f-P_i$ the photon momentum. The relative momenta in the respective terms of 
\Eq{eq:FFeqRL} are defined in Appendix~\ref{sec:kinematics}. The charge matrices 
are defined as
\begin{flalign}\label{eq:charge_matrices}
\mathcal{Q}^{\rho\rho';\lambda}_1&=F^\rho_{abc}\textrm{Q}_{aa'}F^{\rho',\lambda}
_{a'bc} \nonumber\\ 
\mathcal{Q}^{\rho\rho';\lambda}_2&=F^\rho_{abc}\textrm{Q}_{bb'}F^{\rho',\lambda}
_{ab'c} \\
\mathcal{Q}^{\rho\rho';\lambda}_3&=F^\rho_{abc}\textrm{Q}_{cc'}F^{\rho',\lambda}
_{abc'} \nonumber
\end{flalign}
with $\textrm{Q}$ the charge operator
\begin{flalign}
 \textrm{Q}=\left(
     \begin{array}{ccc}
      \nicefrac{2}{3} & 0                & 0 \\
            0         & -\nicefrac{1}{3} & 0 \\      
            0         & 0                & -\nicefrac{1}{3}
     \end{array}\right)~.
\end{flalign}

Finally, the quark-photon vertex $\Gamma^\mu$ is calculated from an
inhomogenous Bethe-Salpeter equation
\begin{flalign}
 \Gamma^\mu(p,Q)&=iZ_2\gamma^\mu\nonumber\\                
&+\int_kK_{q\bar{q}}\left(S(k+Q/2)\Gamma^\mu(k,Q)S(k-Q/2)\right)~,
\end{flalign}
and using for the two-body interaction $K_{q\bar{q}}$ the RL kernel \Eq{eq:RLkernel} with $C=4/3$ and
for the quark propagator $S$ the solutions of the RL-truncated quark DSE. 

We emphasize that 
as a results of this consistent truncation, the quark-photon vertex dynamically develops 
vector-meson poles in its transverse parts \cite{Maris:1999bh,Oettel:2002wf}, 
thus naturally capturing the physics of vector meson dominance
for momenta close to the pole region. This fact has been exploited recently in the context of the pion 
transition form factor leading to good agreement with the experimental data at intermediate momenta and
interesting modifications of the scaling limit at asymptotically large momenta \cite{Eichmann:2017wil}.

%%%%%%%%%%%%%%%%%%%%%%%%%%%%%%%%%%%%%%%%%%%%%%%%%%%%%%%%%%%%%%%%%%%%%
%                        S E C T I O N                              %
%%%%%%%%%%%%%%%%%%%%%%%%%%%%%%%%%%%%%%%%%%%%%%%%%%%%%%%%%%%%%%%%%%%%%

\section{Results}\label{sec:results}

Before we discuss in detail the results for specific transitions, we would like to comment on two technical issues that 
complicate the calculation of form factors in the covariant DSE/BSE approach.

First of all, in the truncation scheme at hand, quark propagators feature complex conjugate poles in the complex momentum plane.
The calculation of transition form factors below or above certain values of $Q^2$ requires probing the quark dressing functions 
in a region which contains poles (see, e.g. \cite{Oettel:2000ig,Eichmann:2007nn,Eichmann:2011aa} 
and also \cite{CPC}). With our current computational 
techniques, calculations are therefore limited to a $Q^2$ window. The specific values of these limits for $Q^2$ depend on 
the masses of the baryons of interest and (where relevant) are indicated in the plots with dashed vertical lines. 
As a matter of fact, in many cases our results are still smooth beyond the high-$Q^2$ limit (see also \cite{Sanchis-Alepuz:2015fcg}) although not so below the low-$Q^2$ limit. 

A second limitation, which becomes critical in the present calculation, stems from the fact that in evaluating \eqref{eq:FFeqRL},
the angles $z_0$, $z_1$ and $z_2$ in \eqref{eq:basis_expansion} become complex and take values beyond a circle of unit radius 
in the complex plane. Since in practice the angular dependence of the Faddeev dressing functions $f(p^2,q^2,z_0,z_1,z_2)$ is
expanded in orthogonal polynomials, such an expansion is strictly non-convergent in this case. It turns out \cite{CPC} that 
this is particularly problematic for transition form factors, except for a small $Q^2$ window (again dependent on the baryon
masses) where the angles are approximately within a unit circle. This allows us to compare the calculation with the angular
dependence truncated in a way to ensure convergence (see e.g. \cite{Eichmann:2011aa}) with the full one. We observed that the
magnetic form factor $G_M$ is reasonably insensitive to such a truncation. This, however, is not the case for the ratios 
$R_{EM}$ and $R_{SM}$. The corresponding results are rather noisy and we chose to show them for the $\gamma^* N\rightarrow \Delta$ transition only; even in this case they should be considered as qualitative only. Note that this 
is not a limitation of the framework itself (since it is formally covariant) but of the numerical implementation of it; 
see \cite{CPC} for a discussion of the necessary steps to overcome such limitation in the future.
 
With these restrictions in mind, we have calculated the electromagnetic transition form factors between the lowest-lying
spin-$\nicefrac{1}{2}$ isospin octet and spin-$\nicefrac{3}{2}$ isospin decuplet, as well as between the neutral octet 
members $\Sigma^0$ and $\Lambda$. In the plots below we present (coloured bands) the numerical results for values of 
the $\eta$ parameter between $\eta=1.6$ (lower bounds) and $\eta=2.0$ (upper bounds).

\subsection{Nucleon - Delta transition}

As already mentioned, the $\gamma^* N\rightarrow \Delta$ transition is the only one with experimental data available
\cite{Bartel:1968tw,Stein:1975yy,Beck:1999ge,Frolov:1998pw,Blanpied:2001ae,Pospischil:2000ad,Sparveris:2004jn,Stave:2008aa,Aznauryan:2009mx}. 
Therefore, most of the theoretical work focused on this transition. This includes lattice QCD
\cite{Alexandrou:2007dt,Alexandrou:2010uk}, effective field theories, large $N_c$ relations, perturbative QCD 
and models, see \cite{Pascalutsa:2006up} for a comprehensive review. It has also been studied previously in the
DSE/BSE approach, using a contact interaction of NJL-type \cite{Segovia:2014aza,Segovia:2016zyc} and in a covariant DSE/BSE 
framework analogous to the present one, i.e. using the same momentum dependent quark-gluon interaction 
Eq.(\ref{eq:MTmodel}) but a quark-diquark approximation for the bound state wave function \cite{Eichmann:2011aa}. For 
the spectrum of ground and 
excited baryon states with light quarks, the quark-diquark approximation has been compared with the three-body approach
very recently \cite{Eichmann:2016hgl} and good agreement has been found. The same is true for the leading parts of the
form factors of the $\Delta^+$, although distinct discrepancies occur for the electric quadrupole and the magnetic octupole
contribution \cite{Nicmorus:2010sd,SanchisAlepuz:2011jn,Eichmann:2016yit}. One would expect similar results for the 
transition form factor.  

The current describing the transition between a spin-$\nicefrac{3}{2}$ and a spin-$\nicefrac{1}{2}$ baryons can be 
written directly in terms of the Jones-Scadron form factors as \cite{Eichmann:2011aa}
\begin{flalign}\label{eq:current_NDG}
 J^{\mu;\rho}\left(P,Q\right)=& i\mathds{P}^{\rho\alpha}\left(P_f\right)\gamma_5\Gamma^{\mu;\alpha}\left(P,Q\right)\Lambda_+\left(P_i\right)~,
\end{flalign}
with $\mathds{P}^{\rho\alpha}$ the Rarita-Schwinger projector for spin-$\nicefrac{3}{2}$ particles and $\Lambda_+$ the positive-energy projectors for Dirac spinors. The vertex $\Gamma$ is defined as
\begin{flalign}\label{eq:vertex_NDg}
\Gamma^{\mu;\alpha}  &= b\left[ \frac{i\omega}{2\lambda_+}\left(G_M^*-G_E^*\right)\gamma_5 \varepsilon^{\alpha\mu\gamma\delta} \widehat{P_T}^\gamma \widehat{Q}^\delta \right.\nonumber \\
                 & \left. -G_E^* T^{\alpha\gamma}_Q T^{\gamma\mu}_{P_T} -\frac{i\tau}{\omega}G_C^* \widehat{Q}^\alpha \widehat{P_T}^\mu \right]~,
\end{flalign}
with 
\begin{flalign}
\tau=& \frac{Q^2}{2\left(M^2_{\nicefrac{3}{2}}+M^2_{\nicefrac{1}{2}}\right)}~,
\hspace*{2mm} \lambda_\pm = \frac{\left(M_{\nicefrac{3}{2}} \pm M_{\nicefrac{1}{2}}\right)^2 + Q^2}{2\left(M_{\nicefrac{3}{2}}^2+M_{\nicefrac{1}{2}}^2\right)}~,\\
\omega= &\sqrt{\lambda_+ \lambda_-}~,
\hspace*{18mm} b = \sqrt{\frac{3}{2}} \left(1 + \frac{M_{\nicefrac{3}{2}}}{M_{\nicefrac{1}{2}}}\right)~,
\end{flalign}
and $T_v$ the transverse projector with respect to the four-vector $v$. Hatted vectors denote unit vectors. The photon momentum is $Q$ and $P=P_f-P_i$, with $P_{i,f}$ the initial and final baryon momenta and $P_T$ is the transverse projection of $P$ with respect to $Q$. Frequently used are the ratios
\begin{flalign}
R_{EM} = -\frac{G^*_{E}}{G^*_M}, \hspace*{10mm} R_{SM}=-\frac{M_N^2}{M_\Delta^2}\sqrt{\lambda_+\lambda_-}\frac{G^*_C}{G^*_M},
\end{flalign}
which at zero momentum are sensitive to the quadrupole moments of the transition.

\begin{figure}[t!]
\begin{center}
\resizebox{0.45\textwidth}{!}{\includegraphics{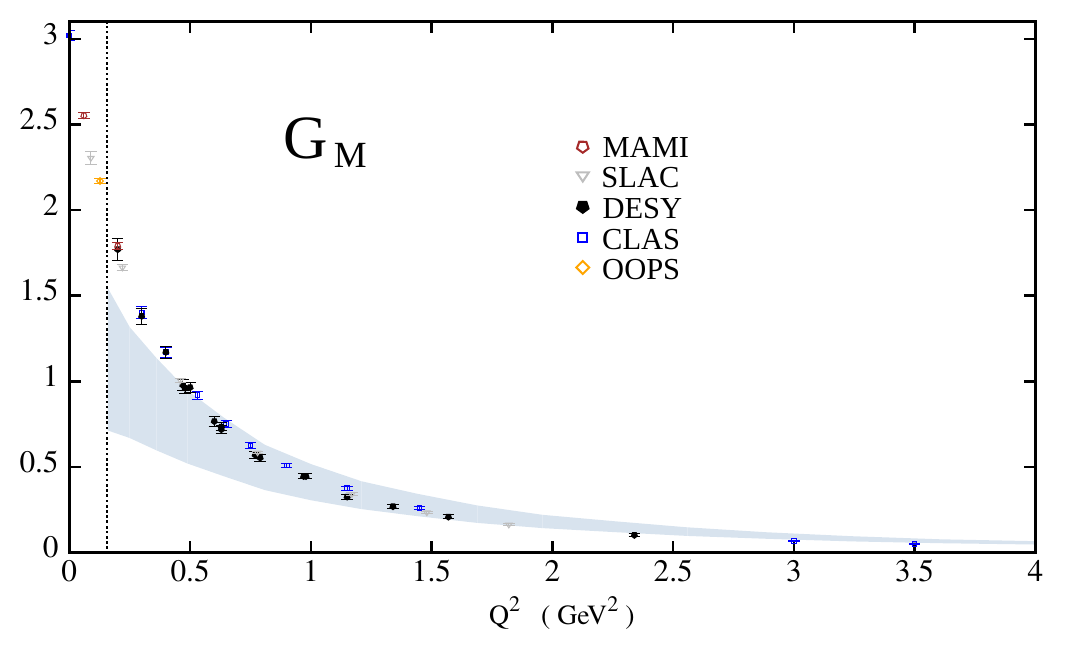}} 
\end{center}
%\caption{Magnetic form factor of the $\gamma^*N\rightarrow\Delta$ transition. Vertical dashed line delimit the region below which the singularities of the quark propagator are probed. Experimental data from \cite{}.
%Coloured bands represent the result of the numerical calculation for $\eta=[1.6,2.0]$.}
%\label{fig:NDg_GM}       
%\end{figure}
%\begin{figure}[t!]
\begin{center}
\resizebox{0.45\textwidth}{!}{\includegraphics{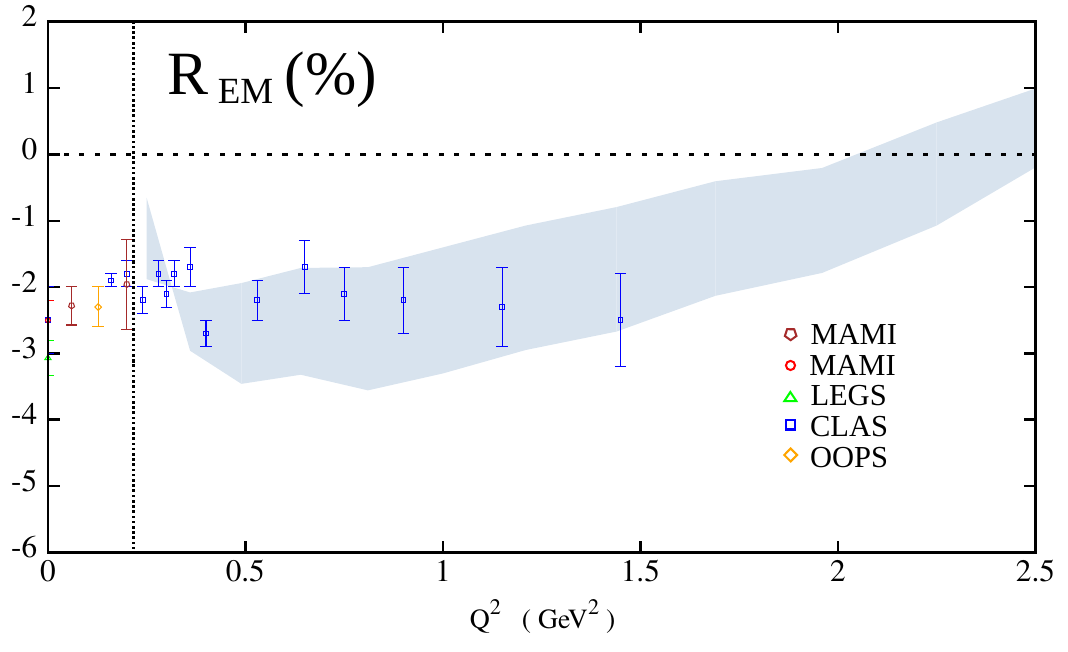}} 
\end{center}
%\caption{Ratio $R_{EM}$ of the $\gamma^*N\rightarrow\Delta$ transition. Vertical dashed line delimit the region below which the singularities of the quark propagator are probed. Experimental data from \cite{}.
%Coloured bands represent the result of the numerical calculation for $\eta=[1.6,2.0]$.}
%\label{fig:NDg_REM}       
%\end{figure}
%\begin{figure}[b!]
\begin{center}
\resizebox{0.45\textwidth}{!}{\includegraphics{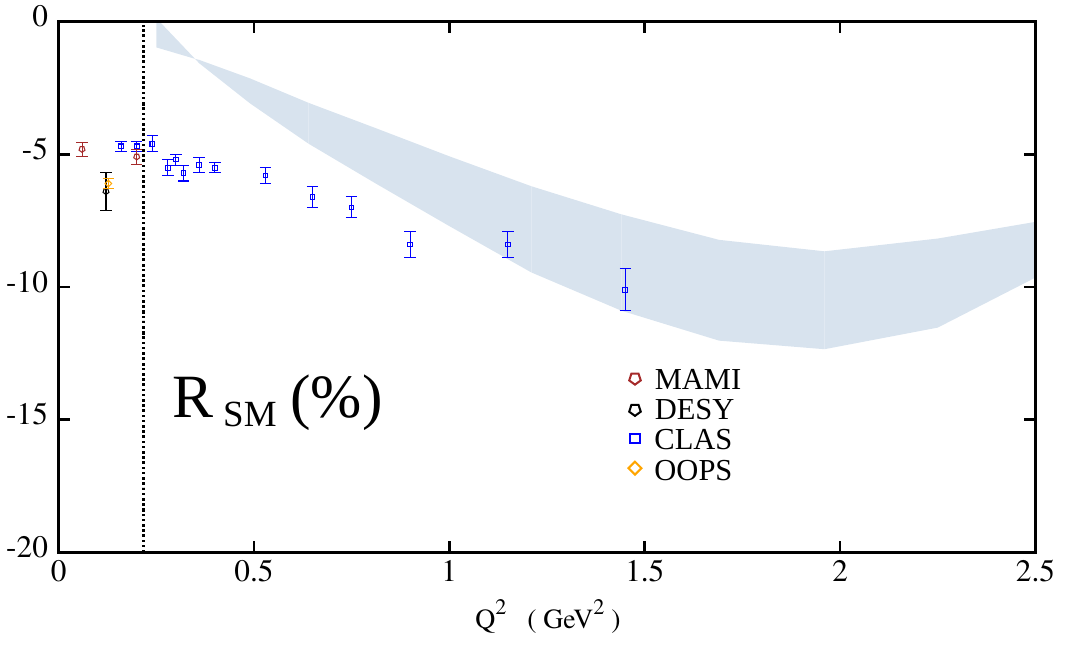}} 
\end{center}
\caption{Magnetic form factor $G_M$, ratio $R_{EM}$ and ratio $R_{SM}$ of the $\gamma^*N\rightarrow\Delta$ transition. 
Vertical dashed line delimit the region below which the singularities of the quark propagator are probed. 
Experimental data are taken from \cite{Bartel:1968tw,Stein:1975yy,Beck:1999ge,Frolov:1998pw,Pospischil:2000ad,Blanpied:2001ae,Sparveris:2004jn,Stave:2008aa,Aznauryan:2009mx}.
Coloured bands represent the result of the numerical calculation for $\eta=[1.6,2.0]$ (lower/upper bound).}
\label{fig:NDg}       
\end{figure}

We show our results in Fig.~\ref{fig:NDg} compared to experimental data. The results for the magnetic form factor are
displayed in the upper panel. Within the theoretical systematic uncertainty estimated by the $\eta$-band our results agree
well with the experimental data for $Q^2 > 0.8$ GeV$^2$. Below we observe deviations leading to a systematic underestimation
of the form factor. This trend is in systematic agreement with others for the nucleon electromagnetic and
axial form factors as well as the form factors for the $\Delta$ \cite{Eichmann:2011vu,Eichmann:2011pv,Sanchis-Alepuz:2013iia}
and indicates the onset of pion cloud effects, which are not included in the present framework (see, however,
\cite{Sanchis-Alepuz:2014wea} for pion cloud corrections to baryon masses). For the ratio $R_{EM}$, displayed in the 
middle panel of Fig.~\ref{fig:NDg}, we confirm an observation
that has been previously discussed in the quark-diquark approximation \cite{Nicmorus:2010sd}. This quantity
is sensitive to the presence of deformations of the nucleon and $\Delta$ due to higher angular momentum. In the non-relativistic
quark model, these come into play via d-wave admixtures to the s-wave structure of the octet baryons. Since, historically,
with d-waves alone the resulting values for $R_{EM}(0)$ were found to be much smaller than the experimental ones, 
it has been concluded that additional pion cloud effects are mandatory. However, as discussed in \cite{Nicmorus:2010sd}, this
is not necessarily true. In a fully covariant framework, as the one at hand, p-wave contributions to the nucleon and the
$\Delta$ are not only allowed, but, them being represented by the lower components of the four-spinors contained in the amplitude, 
and hereby especially also in the leading one, they
 play a much bigger role than d-waves which are related to subleading parts of the baryons' amplitudes. 
Consequently, they contribute to $R_{EM}$ and 
have the potential to generate deviations of $R_{EM}(0)$ from zero of the same order as the experimentally observed ones.
This is true for the diquark-quark approximated framework \cite{Nicmorus:2010sd} and also suggested by our results from
the three-body calculation. This observation also extends to the ratio $R_{SM}(0)$, although here the agreement with 
experiment is much better in the diquark-quark framework than for our results presented in the bottom panel of 
Fig.~\ref{fig:NDg}. In general, as already mentioned above, both ratios do suffer from uncertainties due to technical
problems with the Chebyshev expansion that need to be dealt with in future work. 

\subsection{Hyperon Octet - Decuplet transition}

\begin{figure}%[h!]
\begin{center}
\resizebox{0.39\textwidth}{!}{\includegraphics{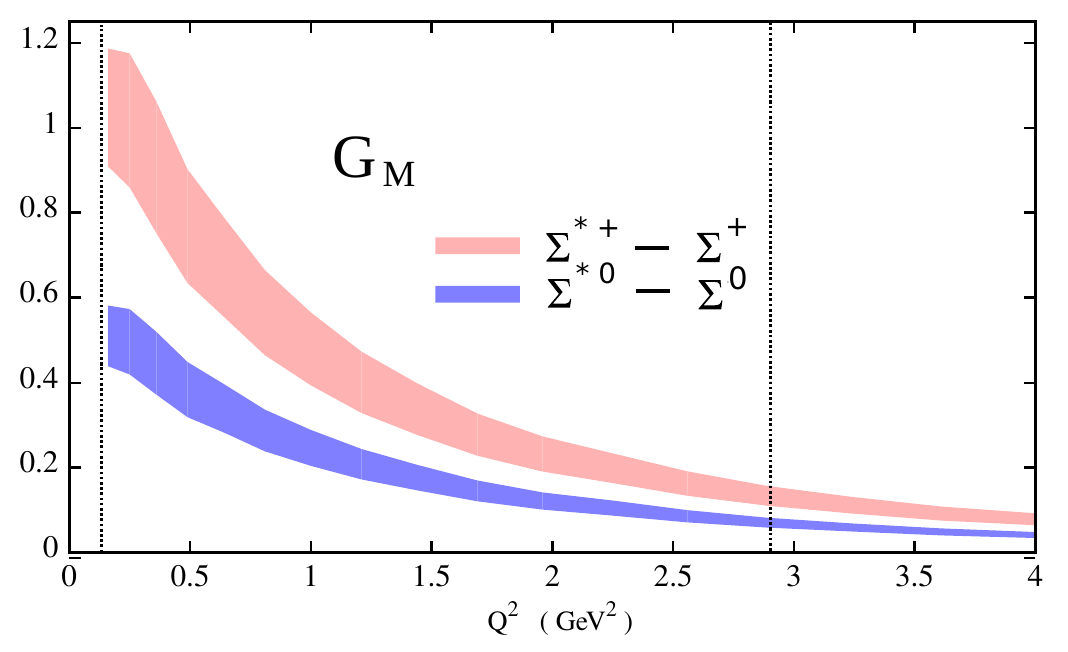}}\\
\resizebox{0.39\textwidth}{!}{\includegraphics{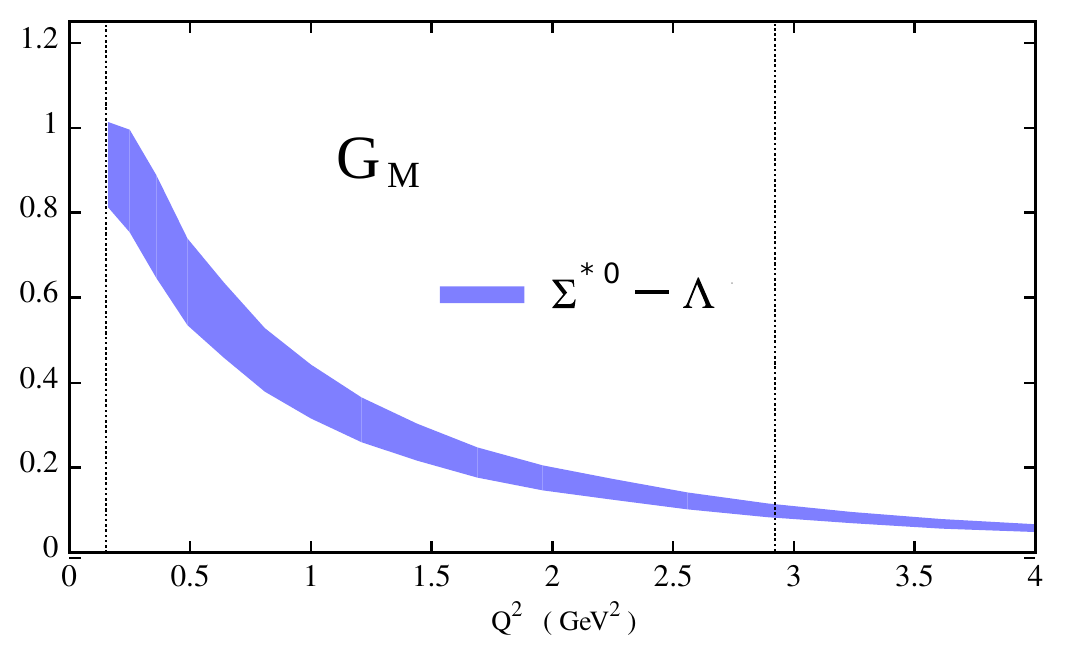}}\\
\resizebox{0.39\textwidth}{!}{\includegraphics{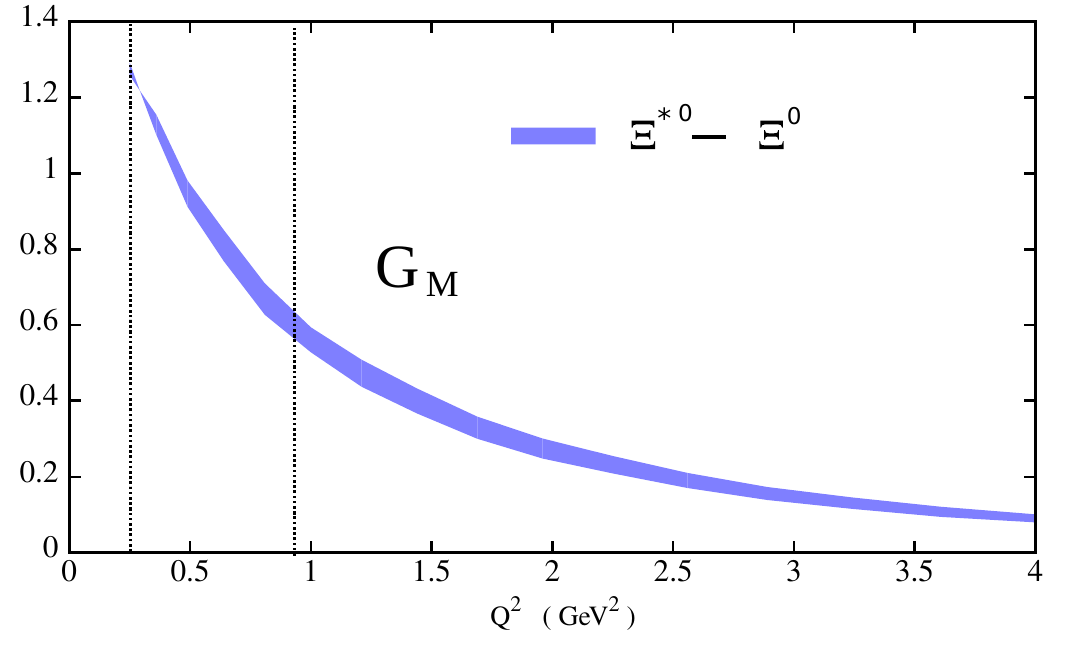}} \\
\resizebox{0.39\textwidth}{!}{\includegraphics{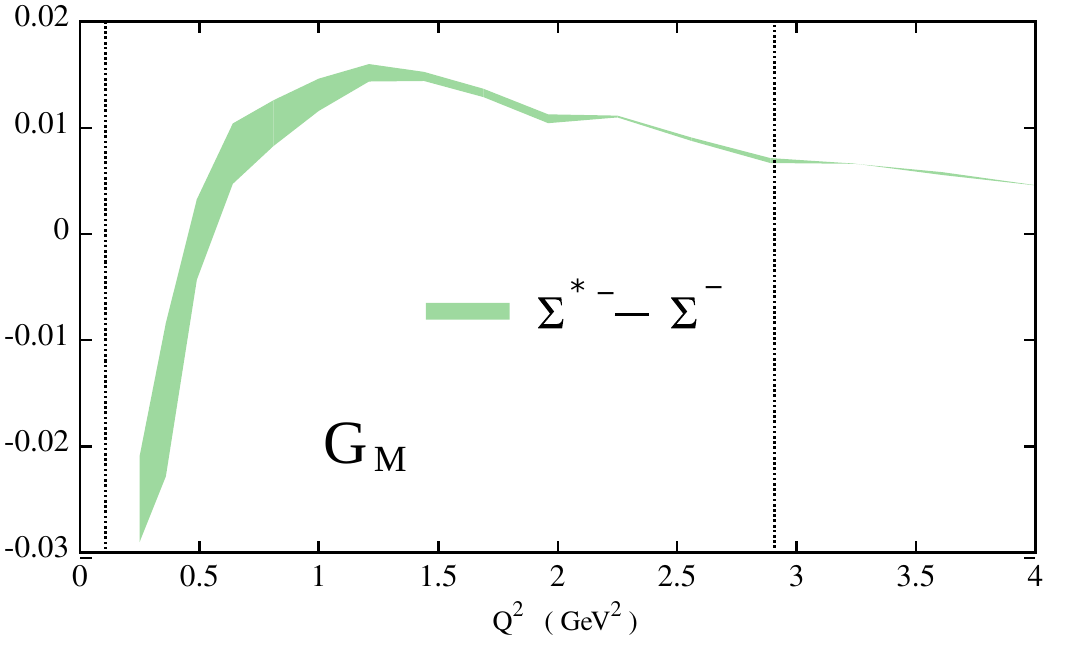}} \\
\resizebox{0.39\textwidth}{!}{\includegraphics{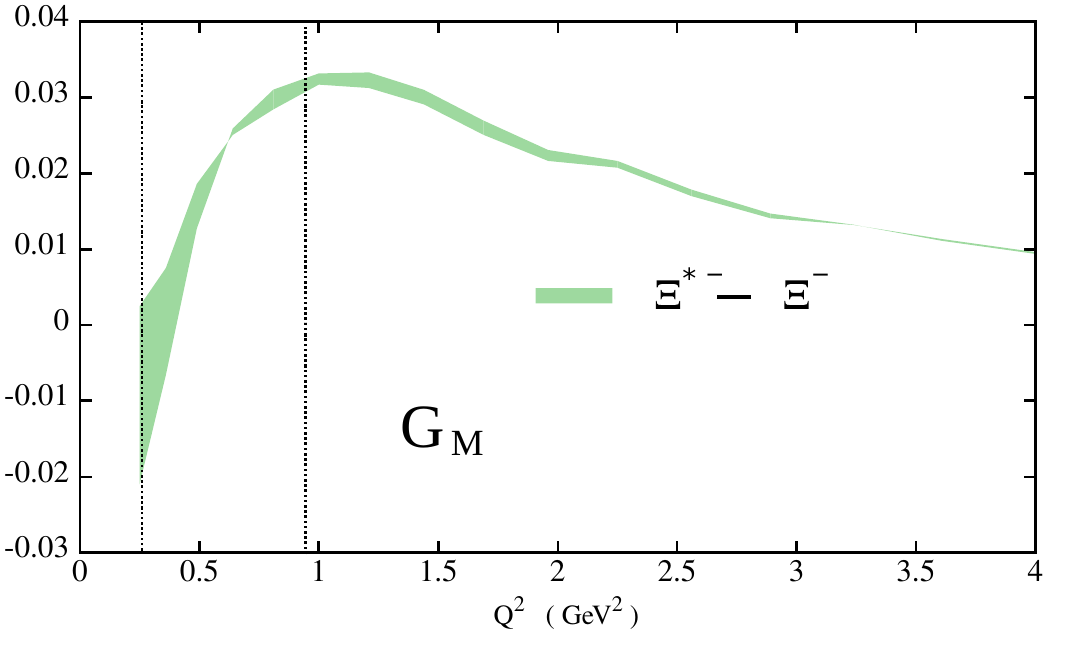}} 
\end{center}
\caption{\label{fig:810} Magnetic form factor of various hyperon transition form factors. 
Dashed vertical lines and coloured bands as in previous figures.}
\end{figure}

\begin{table*}[!t]
\begin{center}
\begin{tabular}{c|ccccccc}
\toprule
transition 		& $\Delta N$	& $\Sigma^{*~+} \Sigma^{+}$ & $\Sigma^{*~0} \Sigma^{0}$ 
                & $\Sigma^{*~0} \Lambda$ & $\Xi^{*~0}\Xi^{*~0}$ & $\Sigma^{*~-} \Sigma^{-}$ & $\Xi^{*~-}\Xi^{*~-}$\\ \midrule 
$G_M(0) (\eta=2.0)$	& 2.0	& 1.1	& 0.5	& 1.0	& 1.8 & -0.05 & -0.07 \\
$G_M(0) (\eta=1.6)$	& 0.8	& 1.3	& 0.6	& 1.1	& 1.5 & -0.04 & -0.02 \\
\hline
\textrm{exp.}	& 3.04~(11) & 4.10~(57) &  & 3.35~(57) &  & $<$0.8 & \\ 
\bottomrule
\end{tabular}
\caption{Extrapolated result for $G_M(0)$ from our quark core calculation and compared to the estimate of the experimental values from \cite{Ramalho:2013uza}.\label{tab:GM0} }
\end{center}
\end{table*}

In the exact $SU(3)$-isospin limit, the transition $\gamma^*\Sigma^{*~+}\rightarrow\Sigma^+$ would be identical to 
the $\gamma^*N\rightarrow\Delta$ studied in the previous section (cf. Eq.~\eqref{eq:current_Sigmap}). This is indeed
manifest in the magnetic form factor, shown in Fig.~\ref{fig:810}, which is comparable in magnitude to the 
corresponding one in Fig.~\ref{fig:NDg} and qualitatively identical in shape. Similar remarks apply to 
the $\gamma^*\Sigma^{*~0}\rightarrow\Sigma^0$, also shown in Fig.~\ref{fig:810}, which is however suppressed 
by a flavour factor (see Eq.~\eqref{eq:current_Sigma0}). 

Once again, in the exact $SU(3)$-isospin limit the $\gamma^*\Xi^{*~0}\rightarrow\Xi^0$ transition would be 
identical to the $N\Delta$ or the $\Sigma^{*+}\Sigma^+$ ones. Comparing the corresponding plots in Fig.~\ref{fig:810} 
we find that their magnetic form factor is indeed very similar. Since the $\Xi$ is a doubly-strange baryon, 
this indicates that the isospin-breaking effects are very small, as we shall see below.

%It is worth mentioning that, in contrast to the $\gamma^*N\rightarrow\Delta$ transition, the form factors in 
%the $\Sigma^{*+}\Sigma^+$, $\Sigma^{*0}\Sigma^0$ and $\Xi^{*~0}\Xi^0$ processes are expected to be less influenced 
%by pion-cloud than kaon-cloud effects \cite{Arndt:2003vd}. Those, however, are suppressed by the kaon mass, 
%the quark core thus providing a good description of the transition. This together with the fact that uncertainties
%due to the Chebychef-expansion are immaterial for $G_M$ would imply that our results could be taken as a quantitative 
%prediction for $G_M(Q^2)$ in these transitions.\cCF{This paragraph needs to be discussed}

A good measure of the breaking of $SU(3)$-flavour symmetry is given by the form factors of the $\gamma^*\Sigma^{*~-}\rightarrow\Sigma^-$ and the $\gamma^*\Xi^{*~-}\rightarrow\Xi^-$ transitions, 
since in the limit of exact symmetry these would vanish identically (cf. Eq.~\eqref{eq:current_Sigmam} and Eq.~\eqref{eq:current_Xim}). We show their magnetic form factors in the two bottom panels in Fig.~\ref{fig:810}. 
They indicate a breaking of $SU(3)$ symmetry at the level of a few percent. The smallness of this breaking 
in the present calculation might result from the fact that it is only generated by the different quark masses 
of the $s-$ and the $u/d-$quarks, both the current quark mass and the dynamically generated one. Other possible 
sources of $SU(3)$ breaking would come, for example, from the weakening of the quark-gluon interaction as a 
function of the quark mass (see \cite{Williams:2014iea}). We should mention, however, that such an effect 
was explored in \cite{Sanchis-Alepuz:2014sca} in relation to the octet and decuplet baryon masses, where 
it was found to be sizeable on the level of the propagators, but extremely small on the level of observables.

We also extrapolated our results for the magnetic form factors to zero momentum using a dipole fit for the 
transitions with positive and neutral charges and a linear extrapolation for the ones with negative charges
applied after the zero crossing. Our results are given in table \ref{tab:GM0} together with the corresponding
results for the nucleon-$\Delta$-transition. As discussed above, we expect significant effects due to meson cloud
effects at small momenta. This is reflected in the sizeable underestimation of $G_M(0)$ as compared with
extractions from experimental values performed in \cite{Ramalho:2013uza}, based on the assumption of the 
dominance of $G_M(0)$ over $G_E(0)$. Compared to the quark model calculation of Ramalho and Tsushima
\cite{Ramalho:2013iaa,Ramalho:2013uza} we indeed find qualitative agreement with their quark core values. 
In their work, corrections due to meson cloud 
effects have been found of the order of 50 percent of the total value for $G_M(0)$. Although such comparisons 
in general need to be taken with a grain of salt (for a start, in any models with parameters the relative sizes 
between quark core and pion cloud contributions are parameter dependent) we regard it as qualitatively indicative 
that the systematic error of our results at small momenta may be of the order of 50 percent. For large momenta, 
however, our results can be taken as a quantitative prediction for $G_M(Q^2)$ in these transitions.

%\cCF{Helios: what did you want to say about zero crossing?}
%\cCF{I think we should show the results for the ratios as well.. the question is: here or in an appendix ?}

\subsection{$\Sigma^0 - \Lambda$ transition}

The $\gamma^*\Sigma^0\rightarrow\Lambda$ transition is the only electromagnetic transition allowed between members 
of the baryon octet. As with other octet hyperon form factors, the experimental information is limited to the magnetic 
moment \cite{Olive:2016xmw}. This reaction, however, has attracted considerable theoretical interest. Its static 
properties have been studied in a number of different models \cite{Kunz:1989me,Park:1990is,Park:1991fb,Wagner:1995va,Aliev:2001uq,Dahiya:2002qj,Cheedket:2002ik,Berdnikov:2007zz,
Sharma:2010vv,Lee:2011nq}, chiral perturbation theory
\cite{Meissner:1997hn,Puglia:1999th,Kubis:2000aa,Geng:2008mf,Geng:2009hh,Ahuatzin:2010ef} 
and quenched lattice QCD \cite{Leinweber:1990dv}. 
The form factors for non-vanishing photon momentum have been studied in \cite{VanCauteren:2005sm,Ramalho:2012ad} 
and very recently using dispersion relations in \cite{Granados:2017cib}. The emphasis of \cite{Ramalho:2012ad} was 
on exploring the role of quark-core and meson-cloud effects on the $\gamma^*\Sigma^0\rightarrow\Lambda$ transition 
form factors. 

\begin{figure}[t!]
\begin{center}
\resizebox{0.45\textwidth}{!}{\includegraphics{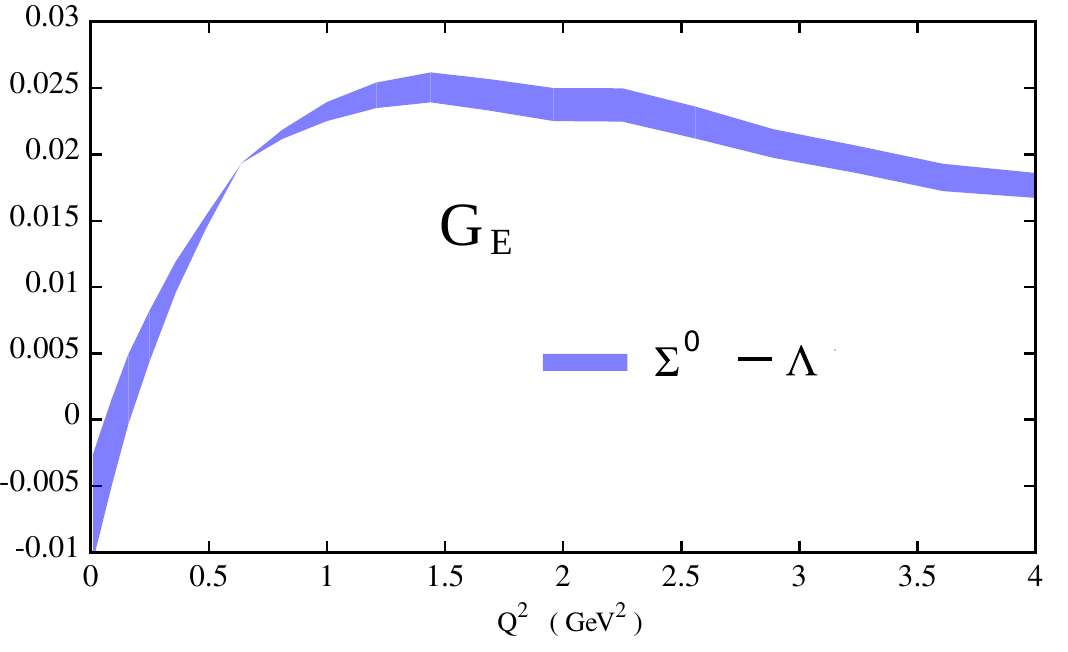}} 
\resizebox{0.44\textwidth}{!}{\includegraphics{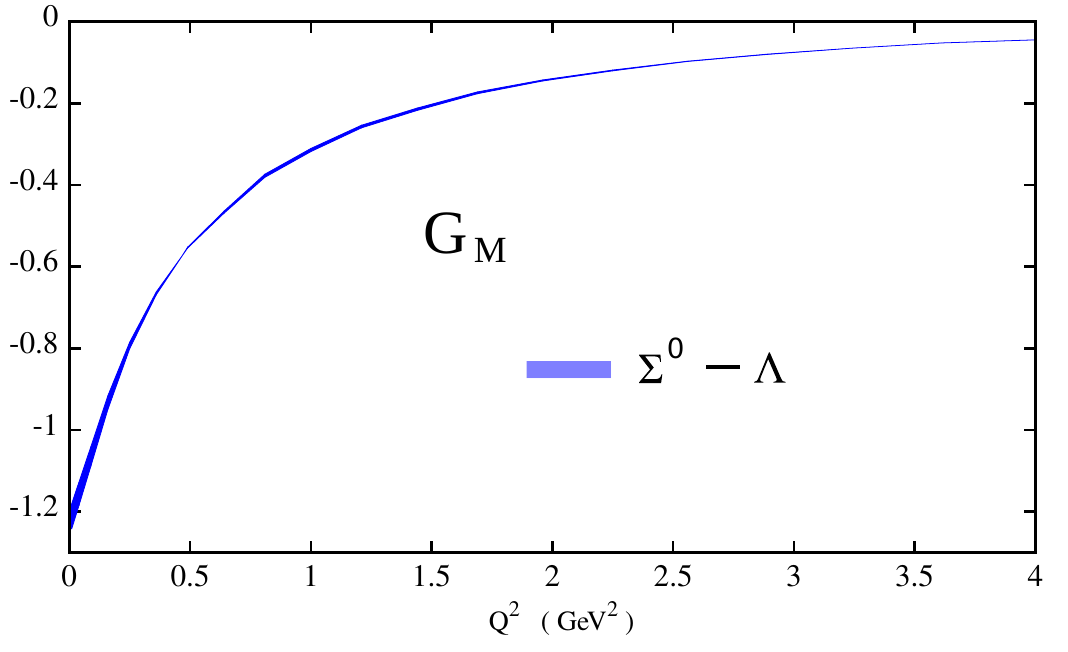}} 
\end{center}
\caption{Electric (upper panel) and magnetic (lower panel) form factors of the 
octet-only $\gamma^*\Sigma^{0}\rightarrow\Lambda$ transition. 
Coloured bands represent the result of the numerical calculation for $\eta=[1.6,2.0]$.}
\label{fig:sigma0_lambda}       
\end{figure}

The form factors for this transition are extracted from the current
\begin{align}\label{eq:SigmaLambda_current}
J^\mu\left(Q,P\right) = i\Lambda_+(P_f) & \left(F_1(Q^2)\left(\gamma^\mu+\frac{\left(M_\Sigma-M_\Lambda\right)}{Q^2}Q^\mu \right)\right.\nonumber\\
                                        & \left.-\frac{F_2(Q^2)}{2\left(M_\Sigma+M_\Lambda\right)}\sigma^{\mu\nu}Q^\nu\right)\Lambda_+(P_i)\,,
\end{align}
with 
$\Lambda^+(\hat{P})=\left(\mathds{1}+\Slash{\hat{P}}\right)/2$ 
the positive-energy projector and the electric and magnetic form factors defined as
\begin{flalign}
 G_E(Q^2)&=F_1(Q^2)-\frac{Q^2}{2 \left(M_\Sigma+M_\Lambda\right)}F_2(Q^2)~,\label{eq:GE_SigmaLambda}\\
 G_M(Q^2)&=F_1(Q^2)+F_2(Q^2)~.
\end{flalign}
In the physical case where the $\Lambda$ and $\Sigma^0$ have different masses, Eq.~\eqref{eq:SigmaLambda_current} 
entails that $F_1$ must vanish for $Q^2=0$ which, from \eqref{eq:GE_SigmaLambda} implies $G_E(0)=0$ as well. 
In our case, however, $\Lambda$ and $\Sigma^0$ are degenerate in mass and this constraint does not apply in 
principle. This is manifest in the upper panel of Fig.~\ref{fig:sigma0_lambda}, where the electric form factor 
is very small but not vanishing at zero photon momentum.

In contrast to the octet-decuplet transitions, our results in Fig.~\ref{fig:sigma0_lambda} show important 
dissimilarities with the quark model results of \cite{Ramalho:2012ad}. There, the quark-core contribution 
to the electric form factor is vanishing, being entirely determined by pion-cloud contributions. This is in 
contrast to our findings, where we obtain a $G_E$ of similar magnitude to that in \cite{Ramalho:2012ad} but 
with meson-cloud effects completely absent. This observation is similar to the one discussed above in connection 
with $R_{EM}$ of the nucleon-$\Delta$ transition and the explanation is the same: In quark models with s-wave contributions
to the wave functions only, these transitions are zero by default. Thus one need to invoke either unnaturally
strong d-wave contributions to the wave-function or attribute the entire non-vanishing form factor to meson 
cloud effects. In our relativistic framework, however, sizeable p-wave contributions to the baryons wave function 
appear naturally and thus generate a non-zero result for $G_E$. 

Another interesting observation is that $G_E$ appears to fall off rather slowly, staying small but 
sizeable for a large $Q^2$-range. The magnetic form factor, shown in Fig.~\ref{fig:sigma0_lambda}, 
shows a trend analogous to other transitions. Judging from the experimental value of the magnetic moment
$|\mu_{\Sigma^0\Lambda}|=1.61~\pm 0.08~\mu_N$ (as well as from experience gained with the calculation of 
elastic hyperon form factors in the same framework \cite{Sanchis-Alepuz:2015fcg}) one can conclude that 
meson-cloud effects play a significant role, at least at low-$Q^2$. Note, finally, that the sign of the 
form factors is not known experimentally. In our calculation, the sign cannot be fixed from this transition 
only, since it can be changed adding a global sign to the $\Lambda$ flavour amplitude (see also the related 
discussion in \cite{Ramalho:2012ad}).

Finally, we extract the slopes of the electric and magnetic form factor at zero momentum. In Ref.~\cite{Granados:2017cib}
a combined approach using dispersion theory and effective field theory has been used to determine for
form factors from small timelike momenta $Q^2=-(m_{\Sigma^0}-m_{\Lambda})^2$ to spacelike momenta of the order
of $Q^2= 1$ GeV$^2$. The results have been found sensitive to uncertainties in the determination of two low-energy
constants, leading to a sizeable spread for the results of the form factors. A determination of the slopes at $Q^2=0$,
either from experiment via precise measurements of the Dalitz-decay $\Sigma^0 \rightarrow \Lambda e^+ e^-$, or from 
theoretical input from other approaches is needed to pin down their results. From our calculation we obtain
\begin{flalign}
\left. \frac{d G_M}{d Q^2}\right|_{Q^2=0} = 1.93..1.75\,, \hspace*{5mm} 
\left. \frac{d G_E}{d Q^2}\right|_{Q^2=0} = 0.053..0.073\,,
\end{flalign}
for the range of values $\eta=[1.6,2.0]$. Naturally these numbers will be modified once meson cloud effects are included.
In the case of the form factors of the nucleon and the $\Sigma^-$, where experimental data are available, the
squared charge radii underestimate the experimental results by 30/25 percent for nucleon/$\Sigma^-$ respectively.
Thus we expect an error of similar size for the derivatives of the transition form factor. 

\section{Summary}\label{sec:summary}

We have presented and discussed results for the electromagnetic transition form factors between all members of the 
baryon octet and decuplet, as well as for the octet-only $\Sigma^0\Lambda$ transition, in the rainbow-ladder truncation 
of the covariant three-body Bethe-Salpeter equation. Such a quark core calculation is generally reliable at
large spacelike photon momentum and, within the $Q^2$-windows indicated in the figures, our results have predictive
power. In the low momentum region, however, our calculation needs to be augmented by the inclusion of meson cloud 
effects. In principle, these effects can be taken into account in a systematic way since their origin 
from the details of the underlying quark-gluon interaction have been clarified in Ref.~\cite{Fischer:2007ze}. 
However, although first calculations of pion cloud effects in the DSE/BSE framework are available for selected 
meson and baryon masses \cite{Fischer:2008wy,Sanchis-Alepuz:2014wea}, their inclusion in form factor calculations 
is a tremendous numerical task, which has not been attempted so far beyond NJL-model type calculations \cite{Cloet:2014rja}.
%\cCF{more refs available ?} 
Without these effects, our results for the form factors at small momentum and the 
associated magnetic moments are too small by up to 50 \%. This is explicitly manifest in the case of the $N\Delta$ 
transition where a comparison with experimental data is available. Similar discrepancies have been observed previously 
in the calculation of strange and non-strange baryon form 
factors \cite{Eichmann:2011vu,Eichmann:2011pv,Sanchis-Alepuz:2013iia,Sanchis-Alepuz:2015fcg}.

An important difference to quark model calculations of transition form factors is the presence of sizeable contributions
of p-wave tensors to the wave function of the baryons. Such terms are inherent to the employed Poincar{\'e}-covariant 
approach: They 
are represented by the lower components of the relativistic four-spinors resulting from the coupling of the three
quarks' Dirac spinors. Especially, also the leading s-wave contribution (upper two components) is always accompanied by a
related p-wave (lower two-components) which are then unavoidable as soon as binding energies are sizeable compared to 
 the related masses. This is different to non- or semi-relativistic quark model
calculations in which the sub-leading d-wave components are held responsible for deviations from the baryons' sphericity. 
For $N\Delta\gamma$ and $\Sigma\Lambda\gamma$ the lower, p-wave, components lead a non-zero
contribution to $G_E$ and the ratio $R_{EM}$ without the need to invoke meson cloud effects. This element is a general
feature of the framework and in principle independent of the truncation. Of course, the precise quantitative contribution
of these p-waves as opposed to meson cloud effects will have to be determined in future calculations.      

\bigskip

\section{Acknowledgements}
This work has been supported by the project P29216-N36
from the Austrian Science Fund, FWF, by the Helmholtz International Center 
for FAIR within the LOEWE program of the State of Hesse, and by the 
DFG collaborative research centre TR 16.

\bigskip \bigskip

%\goodbreak

\newpage

\appendix

\section{Kinematics}
\label{sec:kinematics}

The relative $p,q$ and total $P$ momenta introduced in Eq.~\eqref{eq:BSE_amplitude} are defined in terms of
the three quark momenta $p_1$, $p_2$ and $p_3$ as
\begin{align}\label{eq:defpq}
        p &= (1-\zeta)\,p_3 - \zeta (p_1+p_2)\,, &  p_1 &=  -q -\dfrac{p}{2} +
\dfrac{1-\zeta}{2} P~, \nonumber\\
        q &= \dfrac{p_2-p_1}{2}\,,         &  p_2 &=   q -\dfrac{p}{2} +
\dfrac{1-\zeta}{2} P~,\nonumber\\
        P &= p_1+p_2+p_3\,,                &  p_3 &=   p + \zeta  
P~,
\end{align}
where the momentum partitioning parameter is chosen to be $\zeta=1/3$ . The internal quark 
propagators in the Faddeev equation \Eq{eq:Faddeev_coeff3} depend on 
the internal quark momenta
$k_i=p_i-k$ and $\tilde{k}_i=p_i+k$, with $k$ the exchanged momentum. The 
internal relative momenta, for each of the three terms in the Faddeev equation, are
\begin{align}\label{internal-relative-momenta}
p^{(1)} &= p+k,& p^{(2)} &= p-k,& p^{(3)} &= p,\nonumber\\
q^{(1)} &= q-k/2,& q^{(2)} &= q-k/2, & q^{(3)} &= q+k\,\,.\nonumber\\
\end{align}

The momenta needded for  Eq.~\eqref{eq:Faddeev_coeff} are 
\begin{flalign}
p'=-q-\frac{p}{2}~,\quad q'=-\frac{q}{2}+\frac{3p}{4}~,\nonumber \\
p''=q-\frac{p}{2}~,\quad q''=-\frac{q}{2}-\frac{3p}{4}~.
\end{flalign}
The rotation matrices are
\begin{flalign}\label{eq:rotation_matrices} 
H_1^{ij}&=\left[\bar{\tau}^i_{\beta\alpha\mathcal{I}\gamma}(p,q,P)\tau^j_{
\beta\gamma\alpha\mathcal{I}}(p',q',P)\right]~,\\
H_2^{ij}&=\left[\bar{\tau}^i_{\beta\alpha\mathcal{I}\gamma}(p,q,P)\tau^j_{
\gamma\alpha\beta\mathcal{I}}(p'',q'',P)\right]~.
\end{flalign}

For the calculation of form factors, the $\kappa$-th quark momenta before and after the photon coupling are 
$p_\kappa^{i/f}=p_k\mp Q/2$. The corresponding initial and relative momenta in Eq.~\eqref{eq:FFeqRL} can be then obtained by substituting the respective quark momentum in Eqs.~\eqref{eq:defpq}.

\section{Flavour traces for form factors}\label{sec:flavour}

\begin{table*}[t!]
 \begin{center}
 \small
\renewcommand{\arraystretch}{1.2}
  \begin{tabular}{lcc}\hline
 state 		& A  										&  S   															\\ \hline\hline
$p$  		& $\frac{1}{\sqrt{2}}\left(udu-duu\right)$  & $\frac{1}{\sqrt{6}}\left(2uud-udu-duu\right)$ 				\\[1.0ex]
$n$  		& $\frac{1}{\sqrt{2}}\left(udd-dud\right)$  & $\frac{1}{\sqrt{6}}\left(udd+dud-2ddu\right)$					\\[1.0ex]
$\Sigma^+$ 	& $\frac{1}{\sqrt{2}}\left(usu-suu\right)$ 	& $\frac{1}{\sqrt{6}}\left(2uus-usu-suu\right)$					\\[1.0ex]
$\Sigma^0$ 	& $\frac{1}{2}\left(usd+dsu-sud-sdu\right)$ & $\frac{1}{\sqrt{12}}\left(2uds+2dus-usd-dsu-sud-sdu\right)$ 	\\[1.0ex]
$\Sigma^-$ 	& $\frac{1}{\sqrt{2}}\left(dsd-sdd\right)$ 	& $\frac{1}{\sqrt{6}}\left(2dds-dsd-sdd\right)$ 				\\[1.0ex]
$\Xi^0$ 	& $\frac{1}{\sqrt{2}}\left(uss-sus\right)$ 	& $\frac{1}{\sqrt{6}}\left(uss+sus-2ssu\right)$ 				\\[1.0ex]
$\Xi^-$ 	& $\frac{1}{\sqrt{2}}\left(dss-sds\right)$ 	& $\frac{1}{\sqrt{6}}\left(dss+sds-2ssd\right)$ 				\\ [1.0ex]
$\Lambda^0$ & $\frac{1}{\sqrt{12}}\left(2uds-2dus+sdu-dsu+usd-sud\right)$ & $\frac{1}{2}\left(usd+sud-dsu-sdu\right)$ 	\\[1.0ex]
\hline
\end{tabular}
\caption{Baryon octet flavour amplitudes.\vspace*{5mm}
\label{tab:octet_flavour}}
 \end{center}
\end{table*}

For the calculation of form factors we need the charge matrices $\mathcal{Q}$, 
Eq.~\Eq{eq:charge_matrices} and the matrices
$\mathcal{F}_1$ and $\mathcal{F}_2$, defined as
\begin{align}
 \mathcal{F}^{\rho\rho'}_1=F^\rho_{abc}F^{\rho'}_{bca}~,\\
 \mathcal{F}^{\rho\rho'}_2=F^\rho_{abc}F^{\rho'}_{cab}~.
\end{align}
Using the flavour amplitudes in Tab.~\ref{tab:octet_flavour} and \ref{tab:decuplet_flavour} and denoting the result of the last term in 
\Eq{eq:FFeqRL} for a particular combination of quark flavours 
$(\lambda_1\lambda_2\lambda_3)$ as $J^{\rho}_{\lambda_1\lambda_2\lambda_3}$, 
we get for 
Eq.~\Eq{eq:FFeqRL_simple} in the case of decuplet to octet transition currents (omitting all other indices for clarity),\hfill
\begin{widetext}
\begin{align}
&J_{\Delta^+ p}=J_{\Delta^0 n}=-\sqrt{2} J^{2}_{uuu}~\label{eq:current_NDg},\\
&J_{\Sigma^{*+}\Sigma^+}=-\frac{\sqrt{2}}{3}\left( \sqrt{3}J^{1}_{suu}-\sqrt{3}J^{1}_{usu}+J^{2}_{suu}+J^{2}_{usu}+J^{2}_{uus}\right)\label{eq:current_Sigmap}~,\\
&J_{\Sigma^{*0}\Sigma^0}=-\frac{\sqrt{3}}{6\sqrt{2}}\left( \sqrt{3}J^{1}_{suu}-\sqrt{3}J^{1}_{usu}+J^{2}_{suu}+J^{2}_{usu}+4 J^{2}_{uus}\right)\label{eq:current_Sigma0}~,\\
&J_{\Sigma^{*-}\Sigma^-}=\frac{1}{3\sqrt{2}}\left( \sqrt{3}J^{1}_{suu}-\sqrt{3}J^{1}_{usu}+J^{2}_{suu}+J^{2}_{usu}-2 J^{2}_{uus}\right)\label{eq:current_Sigmam}~,\\
&J_{\Sigma^{*0}\Lambda}=\frac{1}{2\sqrt{2}}\left( J^{1}_{suu}-J^{1}_{usu}-\sqrt{3}\left(J^{2}_{suu}+J^{2}_{usu}\right)\right)\label{eq:current_Sigma32_Lambda}~,\\
&J_{\Xi^{*0}\Xi^0}=\frac{1}{\sqrt{6}}J^{1}_{sus} -\frac{1}{3\sqrt{2}}\left( 
\sqrt{3}J^{1}_{uss}+4 J^{2}_{ssu}+J^{2}_{sus}+J^{2}_{uss}\right)\label{eq:current_Xi0}~,\\
&J_{\Xi^{*-}\Xi^-}=\frac{1}{\sqrt{6}}J^{1}_{sus} -\frac{1}{3\sqrt{2}}\left( 
\sqrt{3}J^{1}_{uss}-2 J^{2}_{ssu}+J^{2}_{sus}+J^{2}_{uss}\right)\label{eq:current_Xim}~,
\end{align}
%\end{widetext}
and for the octet $\Sigma^0 - \Lambda$ transition,\hfill
%\begin{widetext}
\begin{align}
J_{\Sigma^{0}\Lambda}=\frac{1}{4}\left( -\sqrt{3}J^{11}_{suu}-\sqrt{3}J^{11}_{usu}-J^{12}_{suu}+J^{12}_{usu}+3 J^{21}_{suu}-3 J^{21}_{usu}+\sqrt{3}\left( J^{22}_{suu}+J^{22}_{usu}\right)\right)\label{eq:current_SigmaLambda}~.
\end{align}
\end{widetext}

\begin{table}[h!]
 \begin{center}
 \small
\renewcommand{\arraystretch}{1.2}
  \begin{tabular}{lc}\hline
 state 			& S    											\\ \hline\hline
$\Delta^{++}$  	& $uuu$ 										\\[0.7ex]
$\Delta^{+}$  	& $\frac{1}{\sqrt{3}}\left(uud+udu+duu\right)$ 	\\[0.7ex]
$\Delta^{0}$ 	& $\frac{1}{\sqrt{3}}\left(udd+dud+ddu\right)$ 	\\[0.7ex]
$\Delta^{-}$ 	& $ddd$ 										\\[0.7ex]
$\Sigma^{*+}$ 	& $\frac{1}{\sqrt{3}}\left(uus+usu+suu\right)$ 	\\[0.7ex]
$\Sigma^{*0}$ 	& $\frac{1}{\sqrt{6}}\left(uds+usd+dus+dsu+sud+sdu\right)$ \\[0.7ex]
$\Sigma^{*-}$ 	& $\frac{1}{\sqrt{3}}\left(dds+dsd+sdd\right)$ 	\\ [0.7ex]
$\Xi^{*0}$ 		& $\frac{1}{\sqrt{3}}\left(uss+sus+ssu\right)$ 	\\[0.7ex]
$\Xi^{*-}$ 		& $\frac{1}{\sqrt{3}}\left(dss+sds+ssd\right)$ 	\\[0.7ex]
$\Omega^{-}$ 	& $sss$ \\
\hline
\end{tabular}
\caption{Baryon decuplet flavour amplitudes.
\label{tab:decuplet_flavour}}
 \end{center}
\end{table}

\vspace*{30mm}

%
% BibTeX users please use
%\bibliographystyle{spmpsci} 
 \bibliographystyle{epjc}
 \bibliography{hyp_TR_ffs}

\begin{thebibliography}{10}
\providecommand{\url}[1]{\texttt{#1}}
\providecommand{\urlprefix}{URL }
\providecommand{\eprint}[2][]{\url{#2}}

\bibitem{Bernauer:2013tpr}
J.~C. Bernauer, et~al. (A1), Phys. Rev. \textbf{C90}, 1, 015206 (2014),
  \eprint{1307.6227}

\bibitem{Punjabi:2015bba}
V.~Punjabi, C.~F. Perdrisat, M.~K. Jones, et~al., Eur. Phys. J. \textbf{A51},
  79 (2015), \eprint{1503.01452}

\bibitem{Aznauryan:2012ba}
I.~G. Aznauryan, et~al., Int. J. Mod. Phys. \textbf{E22}, 1330015 (2013),
  \eprint{1212.4891}

\bibitem{Olive:2016xmw}
C.~Patrignani, et~al. (Particle Data Group), Chin. Phys. \textbf{C40}, 10,
  100001 (2016)

\bibitem{Dobbs:2014ifa}
S.~Dobbs, A.~Tomaradze, T.~Xiao, et~al., Phys. Lett. \textbf{B739}, 90 (2014),
  \eprint{1410.8356}

\bibitem{Keller:2011nt}
D.~Keller, et~al. (CLAS), Phys. Rev. \textbf{D83}, 072004 (2011),
  \eprint{1103.5701}

\bibitem{Keller:2011aw}
D.~Keller, et~al. (CLAS), Phys. Rev. \textbf{D85}, 052004 (2012),
  \eprint{1111.5444}

\bibitem{Dudek:2012vr}
J.~Dudek, et~al., Eur. Phys. J. \textbf{A48}, 187 (2012), \eprint{1208.1244}

\bibitem{Granados:2017cib}
C.~Granados, S.~Leupold, E.~Perotti, Eur. Phys. J. \textbf{A53}, 6, 117 (2017),
  \eprint{1701.09130}

\bibitem{Eichmann:2009qa}
G.~Eichmann, R.~Alkofer, A.~Krassnigg, et~al., Phys. Rev. Lett. \textbf{104},
  201601 (2010), \eprint{0912.2246}

\bibitem{Eichmann:2011vu}
G.~Eichmann, Phys. Rev. \textbf{D84}, 014014 (2011), \eprint{1104.4505}

\bibitem{Eichmann:2011pv}
G.~Eichmann, C.~S. Fischer, Eur. Phys. J. \textbf{A48}, 9 (2012),
  \eprint{1111.2614}

\bibitem{Sanchis-Alepuz:2013iia}
H.~Sanchis-Alepuz, R.~Williams, R.~Alkofer, Phys. Rev. \textbf{D87}, 9, 096015
  (2013), \eprint{1302.6048}

\bibitem{Sanchis-Alepuz:2015fcg}
H.~Sanchis-Alepuz, C.~S. Fischer, Eur. Phys. J. \textbf{A52}, 2, 34 (2016),
  \eprint{1512.00833}

\bibitem{Eichmann:2016hgl}
G.~Eichmann, C.~S. Fischer, H.~Sanchis-Alepuz, Phys. Rev. \textbf{D94}, 9,
  094033 (2016), \eprint{1607.05748}

\bibitem{Eichmann:2016yit}
G.~Eichmann, H.~Sanchis-Alepuz, R.~Williams, et~al., Prog. Part. Nucl. Phys.
  \textbf{91}, 1 (2016), \eprint{1606.09602}

\bibitem{Eichmann:2009en}
G.~Eichmann, R.~Alkofer, A.~Krassnigg, et~al., EPJ Web Conf. \textbf{3}, 03028
  (2010), \eprint{0912.2876}

\bibitem{Eichmann:2009zx}
G.~Eichmann, \emph{{Hadron Properties from QCD Bound-State Equations}}, Ph.D.
  thesis, Graz U. (2009), \eprint{0909.0703},
  \urlprefix\url{https://inspirehep.net/record/830292/files/arXiv:0909.0703.pdf}

\bibitem{SanchisAlepuz:2011jn}
H.~Sanchis-Alepuz, G.~Eichmann, S.~Villalba-Chavez, et~al., Phys. Rev.
  \textbf{D84}, 096003 (2011), \eprint{1109.0199}

\bibitem{Maris:1997tm}
P.~Maris, C.~D. Roberts, Phys. Rev. \textbf{C56}, 3369 (1997),
  \eprint{nucl-th/9708029}

\bibitem{Maris:1999nt}
P.~Maris, P.~C. Tandy, Phys. Rev. \textbf{C60}, 055214 (1999),
  \eprint{nucl-th/9905056}

\bibitem{Krassnigg:2009zh}
A.~Krassnigg, Phys. Rev. \textbf{D80}, 114010 (2009), \eprint{0909.4016}

\bibitem{Nicmorus:2010mc}
D.~Nicmorus, G.~Eichmann, A.~Krassnigg, et~al., Few Body Syst. \textbf{49}, 255
  (2011), \eprint{1008.4149}

\bibitem{Haberzettl:1997jg}
H.~Haberzettl, Phys. Rev. \textbf{C56}, 2041 (1997), \eprint{nucl-th/9704057}

\bibitem{Kvinikhidze:1998xn}
A.~N. Kvinikhidze, B.~Blankleider, Phys. Rev. \textbf{C60}, 044003 (1999),
  \eprint{nucl-th/9901001}

\bibitem{Kvinikhidze:1999xp}
A.~N. Kvinikhidze, B.~Blankleider, Phys. Rev. \textbf{C60}, 044004 (1999),
  \eprint{nucl-th/9901002}

\bibitem{Oettel:1999gc}
M.~Oettel, M.~Pichowsky, L.~von Smekal, Eur. Phys. J. \textbf{A8}, 251 (2000),
  \eprint{nucl-th/9909082}

\bibitem{Oettel:2000jj}
M.~Oettel, R.~Alkofer, L.~von Smekal, Eur. Phys. J. \textbf{A8}, 553 (2000),
  \eprint{nucl-th/0006082}

\bibitem{Maris:1999bh}
P.~Maris, P.~C. Tandy, Phys. Rev. \textbf{C61}, 045202 (2000),
  \eprint{nucl-th/9910033}

\bibitem{Oettel:2002wf}
M.~Oettel, R.~Alkofer, Eur. Phys. J. \textbf{A16}, 95 (2003),
  \eprint{hep-ph/0204178}

\bibitem{Eichmann:2017wil}
G.~Eichmann, C.~Fischer, E.~Weil, et~al.  (2017), \eprint{1704.05774}

\bibitem{Oettel:2000ig}
M.~Oettel, \emph{{Baryons as relativistic bound states of quark and diquark}},
  Ph.D. thesis, Tubingen U. (2000), \eprint{nucl-th/0012067}

\bibitem{Eichmann:2007nn}
G.~Eichmann, A.~Krassnigg, M.~Schwinzerl, et~al., Annals Phys. \textbf{323},
  2505 (2008), \eprint{0712.2666}

\bibitem{Eichmann:2011aa}
G.~Eichmann, D.~Nicmorus, Phys. Rev. \textbf{D85}, 093004 (2012),
  \eprint{1112.2232}

\bibitem{CPC}
H.~Sanchis-Alepuz, R.~Williams, Comp. Phys. Comm. (in preparation)  (2017)

\bibitem{Bartel:1968tw}
W.~Bartel, B.~Dudelzak, H.~Krehbiel, et~al., Phys. Lett. \textbf{B28}, 148
  (1968)

\bibitem{Stein:1975yy}
S.~Stein, W.~B. Atwood, E.~D. Bloom, et~al., Phys. Rev. \textbf{D12}, 1884
  (1975)

\bibitem{Beck:1999ge}
R.~Beck, et~al., Phys. Rev. \textbf{C61}, 035204 (2000),
  \eprint{nucl-ex/9908017}

\bibitem{Frolov:1998pw}
V.~V. Frolov, et~al., Phys. Rev. Lett. \textbf{82}, 45 (1999),
  \eprint{hep-ex/9808024}

\bibitem{Blanpied:2001ae}
G.~Blanpied, et~al., Phys. Rev. \textbf{C64}, 025203 (2001)

\bibitem{Pospischil:2000ad}
T.~Pospischil, et~al., Phys. Rev. Lett. \textbf{86}, 2959 (2001),
  \eprint{nucl-ex/0010020}

\bibitem{Sparveris:2004jn}
N.~F. Sparveris, et~al. (OOPS), Phys. Rev. Lett. \textbf{94}, 022003 (2005),
  \eprint{nucl-ex/0408003}

\bibitem{Stave:2008aa}
S.~Stave, et~al. (A1), Phys. Rev. \textbf{C78}, 025209 (2008),
  \eprint{0803.2476}

\bibitem{Aznauryan:2009mx}
I.~G. Aznauryan, et~al. (CLAS), Phys. Rev. \textbf{C80}, 055203 (2009),
  \eprint{0909.2349}

\bibitem{Alexandrou:2007dt}
C.~Alexandrou, G.~Koutsou, H.~Neff, et~al., Phys. Rev. \textbf{D77}, 085012
  (2008), \eprint{0710.4621}

\bibitem{Alexandrou:2010uk}
C.~Alexandrou, G.~Koutsou, J.~W. Negele, et~al., Phys. Rev. \textbf{D83},
  014501 (2011), \eprint{1011.3233}

\bibitem{Pascalutsa:2006up}
V.~Pascalutsa, M.~Vanderhaeghen, S.~N. Yang, Phys. Rept. \textbf{437}, 125
  (2007), \eprint{hep-ph/0609004}

\bibitem{Segovia:2014aza}
J.~Segovia, I.~C. Cloet, C.~D. Roberts, et~al., Few Body Syst. \textbf{55},
  1185 (2014), \eprint{1408.2919}

\bibitem{Segovia:2016zyc}
J.~Segovia, C.~D. Roberts, Phys. Rev. \textbf{C94}, 4, 042201 (2016),
  \eprint{1607.04405}

\bibitem{Nicmorus:2010sd}
D.~Nicmorus, G.~Eichmann, R.~Alkofer, Phys. Rev. \textbf{D82}, 114017 (2010),
  \eprint{1008.3184}

\bibitem{Sanchis-Alepuz:2014wea}
H.~Sanchis-Alepuz, C.~S. Fischer, S.~Kubrak, Phys. Lett. \textbf{B733}, 151
  (2014), \eprint{1401.3183}

\bibitem{Ramalho:2013uza}
G.~Ramalho, K.~Tsushima, Phys. Rev. \textbf{D87}, 9, 093011 (2013),
  \eprint{1302.6889}

\bibitem{Williams:2014iea}
R.~Williams, Eur. Phys. J. \textbf{A51}, 5, 57 (2015), \eprint{1404.2545}

\bibitem{Sanchis-Alepuz:2014sca}
H.~Sanchis-Alepuz, C.~S. Fischer, Phys. Rev. \textbf{D90}, 9, 096001 (2014),
  \eprint{1408.5577}

\bibitem{Ramalho:2013iaa}
G.~Ramalho, K.~Tsushima, Phys. Rev. \textbf{D88}, 053002 (2013),
  \eprint{1307.6840}

\bibitem{Kunz:1989me}
J.~Kunz, P.~J. Mulders, Phys. Rev. \textbf{D41}, 1578 (1990)

\bibitem{Park:1990is}
N.~W. Park, J.~Schechter, H.~Weigel, Phys. Rev. \textbf{D43}, 869 (1991)

\bibitem{Park:1991fb}
N.~W. Park, H.~Weigel, Nucl. Phys. \textbf{A541}, 453 (1992)

\bibitem{Wagner:1995va}
G.~Wagner, A.~J. Buchmann, A.~Faessler, Phys. Lett. \textbf{B359}, 288 (1995),
  \eprint{nucl-th/9507032}

\bibitem{Aliev:2001uq}
T.~M. Aliev, A.~Ozpineci, M.~Savci, Phys. Lett. \textbf{B516}, 299 (2001),
  \eprint{hep-ph/0104204}

\bibitem{Dahiya:2002qj}
H.~Dahiya, M.~Gupta, Phys. Rev. \textbf{D67}, 114015 (2003),
  \eprint{hep-ph/0211127}

\bibitem{Cheedket:2002ik}
S.~Cheedket, V.~E. Lyubovitskij, T.~Gutsche, et~al., Eur. Phys. J.
  \textbf{A20}, 317 (2004), \eprint{hep-ph/0212347}

\bibitem{Berdnikov:2007zz}
A.~{\relax Ya}. Berdnikov, {\relax Ya}.~A. Berdnikov, A.~N. Ivanov, et~al.,
  Mod. Phys. Lett. \textbf{A22}, 2265 (2007)

\bibitem{Sharma:2010vv}
N.~Sharma, H.~Dahiya, P.~K. Chatley, et~al., Phys. Rev. \textbf{D81}, 073001
  (2010), \eprint{1003.4338}

\bibitem{Lee:2011nq}
F.~X. Lee, L.~Wang, Phys. Rev. \textbf{D83}, 094008 (2011), \eprint{1102.1710}

\bibitem{Meissner:1997hn}
U.-G. Meissner, S.~Steininger, Nucl. Phys. \textbf{B499}, 349 (1997),
  \eprint{hep-ph/9701260}

\bibitem{Puglia:1999th}
S.~J. Puglia, M.~J. Ramsey-Musolf, Phys. Rev. \textbf{D62}, 034010 (2000),
  \eprint{hep-ph/9911542}

\bibitem{Kubis:2000aa}
B.~Kubis, U.~G. Meissner, Eur. Phys. J. \textbf{C18}, 747 (2001),
  \eprint{hep-ph/0010283}

\bibitem{Geng:2008mf}
L.~S. Geng, J.~Martin~Camalich, L.~Alvarez-Ruso, et~al., Phys. Rev. Lett.
  \textbf{101}, 222002 (2008), \eprint{0805.1419}

\bibitem{Geng:2009hh}
L.~S. Geng, J.~Martin~Camalich, M.~J. Vicente~Vacas, Phys. Lett. \textbf{B676},
  63 (2009), \eprint{0903.0779}

\bibitem{Ahuatzin:2010ef}
G.~Ahuatzin, R.~Flores-Mendieta, M.~A. Hernandez-Ruiz, Phys. Rev. \textbf{D89},
  3, 034012 (2014), \eprint{1011.5268}

\bibitem{Leinweber:1990dv}
D.~B. Leinweber, R.~M. Woloshyn, T.~Draper, Phys. Rev. \textbf{D43}, 1659
  (1991)

\bibitem{VanCauteren:2005sm}
T.~Van~Cauteren, J.~Ryckebusch, B.~Metsch, et~al., Eur. Phys. J. \textbf{A26},
  339 (2005), \eprint{nucl-th/0509047}

\bibitem{Ramalho:2012ad}
G.~Ramalho, K.~Tsushima, Phys. Rev. \textbf{D86}, 114030 (2012),
  \eprint{1210.7465}

\bibitem{Fischer:2007ze}
C.~S. Fischer, D.~Nickel, J.~Wambach, Phys. Rev. \textbf{D76}, 094009 (2007),
  \eprint{0705.4407}

\bibitem{Fischer:2008wy}
C.~S. Fischer, R.~Williams, Phys. Rev. \textbf{D78}, 074006 (2008),
  \eprint{0808.3372}

\bibitem{Cloet:2014rja}
I.~C. Cloët, W.~Bentz, A.~W. Thomas, Phys. Rev. \textbf{C90}, 045202 (2014),
  \eprint{1405.5542}

\end{thebibliography}
%\begin{thebibliography}{}
%

%\end{thebibliography}

\end{document}